# Time-domain extreme ultraviolet diffuse scattering spectroscopy of nanoscale surface phonons

F. Capotondi[1,*,†], A.A. Maznev[2,*,‡], F. Bencivenga[1,*], S. Bonetti[3], D. Engel[4], D. Fainozzi[1,5], D. Fausti[1,6,7], L. Foglia[1], C. Gutt[8], N. Jaouen[9,3], D. Ksenzov[8], C. Masciovecchio[1], Keith A. Nelson[2], I. Nikolov[1], M. Pancaldi[1,3], E. Pedersoli[1], B. Pfau[4], L. Raimondi[1,10], F. Romanelli[11], R. Totani[1], M. Trigo[12,13]

*[1]Elettra Sincrotrone Trieste, Strada Statale 14, km 163.5, 34149 Basovizza, TS, Italy.*
*[2]Department of Chemistry, Massachusetts Institute of Technology, Cambridge, Massachusetts 02139, USA.*
*[3] Department of Molecular Sciences and Nanosystems, Ca' Foscari University of Venice, Venice, Italy.*
*[4] Max Born Institute, Max-Born-Straße 2A, 12489 Berlin, Germany.*
*[5] Institute of Applied Physics, University of Bern, Sidlerstrasse 5, CH-3012 Bern, Switzerland.*
*[6]Department of Physics, Università degli Studi di Trieste, 34127 Trieste, Italy.*
*[7]Department of Physics, University of Erlangen-Nürnberg, 91058 Erlangen, Germany.*
*[8]Department of Physics, University of Siegen, Walter-Flex-Strasse 3, 57072 Siegen, Germany.*
*[9] Synchrotron SOLEIL, Saint-Aubin, Boite Postale 48, 91192, Gif-sur-Yvette Cedex, France.*
*[10] Advanced Light Source, Lawrence Berkeley National Lab, Berkeley, CA 94720, USA.*
*[11] Department of Mathematics, Informatics and Geosciences, Università degli Studi di Trieste, 34128 Trieste, Italy.*
*[12] Stanford Institute for Materials and Energy Sciences, SLAC National Accelerator Laboratory, Menlo Park, California 94025, USA.*
*[13] Stanford PULSE Institute, SLAC National Accelerator Laboratory, Menlo Park, California 94025, USA.*

** These authors contributed equally to this work.*
*† Contact author: flavio.capotondi@elettra.eu*
*‡ Contact author: alexei.maznev@gmail.com*

We report the observation of dynamic fringe patterns in the diffuse scattering of extreme ultraviolet light from surfaces, following femtosecond optical excitation. At each point on the detector, the diffuse scattering intensity exhibits oscillations at well-defined frequencies that correspond to surface phonons with wave vectors determined by the scattering geometry, indicating that the optical excitation generates coherent surface phonons propagating in all directions and spanning a wavelength range from 60 to 300 nm. This phenomenon is observed on a variety of samples, including single-layer and multilayer metal films, as well as bulk semiconductors. The measured surface phonon dispersions show good agreement with theoretical calculations. By comparing signal amplitudes from samples with different surface morphologies, we find that the excitation mechanism is linked to the natural surface roughness of the samples. However, the signal is still detectable on extremely smooth surfaces with sub-nanometer roughness. Our findings demonstrate a simple and effective method for optically exciting coherent surface phonons with nanoscale wavelengths on a wide range of solid samples and establish a foundation for surface phonon spectroscopy in a wave vector range well beyond the limit of conventional surface Brillouin scattering.

Surface acoustic waves (SAWs) propagating at solid surfaces play an important role in a wide range of fields, from seismology to SAW filtering technologies used in telecommunications devices. The interaction of light with the thermal population of surface acoustic phonons gives rise to surface Brillouin scattering (SBS) [1,2]. However, the wave vector of phonons accessible via SBS is limited to twice the optical wave vector, therefore SBS conducted with optical wavelengths cannot detect phonons with wavelengths below a few hundred nanometers. Scattering of extreme ultraviolet (EUV) light would, in principle, allow access to shorter wavelength phonons. However, SBS in the EUV has been hampered by the lack of high-resolution spectrometers. Recently, the transient grating (TG) approach has been extended into the EUV range [3,4], enabling the excitation and detection of coherent SAWs at sub-100 nm wavelengths [5]. Alternatively, nanoscale SAWs can be generated by optical laser excitation of nanostructures fabricated on the sample surface [6,7]. In both approaches, the generation of short-wavelength coherent SAWs relies on patterning, either of the excitation light, as in the TG method, or of the sample surface. Without such patterning, optical excitation normally generates SAWs with long wavelengths on the order of the optical spot size [8].

In this letter, we report an unexpected observation of diffuse EUV scattering (DES) from spatially random, but temporally coherent, nanoscale SAWs excited by uniform femtosecond optical laser irradiation of an unpatterned sample surface. Following the excitation, DES produces a distinct circular fringe pattern evolving with the time delay $t$. At each point on a bidimensional detector, DES intensity oscillates at a frequency matching the SAW frequency at the wave vector defined by the scattering geometry. This phenomenon is observed across a variety of samples indicating a universal excitation mechanism, applicable to any material that absorbs the pump radiation. By comparing samples with different surface morphologies, we identify the natural surface roughness as the factor enabling the excitation of nanoscale SAWs.

The experiment was conducted at the DiProI end station at the FERMI free electron laser (FEL) facility [9]. The optical (pump) and FEL (probe) pulses are incident on the sample surface at 45° (Fig. 1(a)). DES of a nearly transform limited FEL pulse ($\lambda_{FEL}$ = 17.8 nm, pulse duration 50 fs, focused to a spot size of ~150x120 μm$^2$ with a fluence of ~1.5 mJ/cm$^2$) is recorded by a CCD detector positioned 50 mm from the interaction point. To trigger the lattice dynamics, an optical laser pulse ($\lambda_{opt}$ = 395 nm, pulse duration 80 fs) is focused on the same location, delivering a fluence between 7.5 and 30 mJ/cm$^2$ into a spot of ~300x250 μm$^2$ in size. Figure 1(b) shows a featureless static DES intensity pattern, $I_{\mathrm{DES}}^{Static}(Q_x, Q_y)$, where $Q_x$ and $Q_y$ represent the components of the scattering wave vector, from a [Pt(4 nm)/Al(4 nm)]×4 multilayer stack deposited on Si/$SiO_2$(280 nm) substrate. The detector was exposed to 600 FEL shots at 50 Hz repetition rate. The excitation induces significant variations – up to a few percent – in the DES intensity $I_{\mathrm{DES}}^{Pumped}(Q_x, Q_y, t)$, which are clearly observable in the differential image $\Delta I_{\mathrm{DES}}(Q_x, Q_y, t) = I_{\mathrm{DES}}^{Pumped}(Q_x, Q_y, t) - I_{\mathrm{DES}}^{Static}(Q_x, Q_y)$. Figures 1(c)-(f) show snapshots of $\Delta I_{\mathrm{DES}}$ at four representative delays. While for a negative delay, $\Delta I_{\mathrm{DES}}$ appears featureless, for positive delays, a time dependent concentric ring pattern is clearly observable at each time step. A movie showing the time evolution of this fringe pattern is available in Supplemental Movie 1 [10]. Figure 2(a) presents the time evolution of the radial average of the differential DES intensity, defined as:

$$S(Q, t) = \int_0^{2\pi} \Delta I_{\mathrm{DES}}(Q, \phi, t) d\phi, \qquad (1)$$

where $Q$ and $\phi$ are the magnitude and angle of the scattering wave vector. Figure 2(b) shows that at a given $Q$, the DES intensity oscillates at a frequency that increases with $Q$.

In Figure 2(c), we present the temporal Fourier transform of the data reported in Figure 2(a), showing the dependence of the oscillation frequency on the wave vector $Q$. The experimental data closely match the calculated dispersion curve (Supplemental Section S5 [10]) for the surface Rayleigh mode of the sample [12,13]. Figure 2(d) reveals a linear dependence of the modulation amplitude of $S(Q, t)$ at $Q$ = 0.05 nm$^{-1}$ on the optical laser fluence $F_{ex}$, in the range 7.5 – 30 mJ/cm$^2$. The linear dependence is remarkable, as in both SBS and TG experiments the scattering intensity is proportional to the square of the surface displacement [2,5]. As will be shown below, this dependence is explained by the interference of the scattering from laser-excited SAWs with static scattering from surface roughness.

The association of the observed phenomenon with SAWs is further supported by measurements on additional samples. Figures 3(a) and 3(b) show differential DES images at $t$ = 200 ps for a bulk (001) GaAs sample and a 100 nm thick Ti film on a Si substrate, respectively. Figures 3(c) and 3(d) display the temporal Fourier transforms of the corresponding $S(Q, t)$ maps. On GaAs, the signal is lower compared to metal film samples, but clear oscillations are observed across the entire probed $Q$-range. Figure 3(c) reveals a linear dispersion with a slope of 3050±50 m/s, which is in line with the average Rayleigh velocity for (001) GaAs [14]. A further investigation reveals a slight anisotropy: as expected [14], the SAW velocity is larger for the [110] propagation direction compared to the [100] direction, see Supplemental Section S2 [10]. For the Ti film on Si, the Fourier transform reveals the presence of multiple modes, with a good agreement between the measured and calculated dispersion [12,13]. Measurements on a number of other samples (Supplemental Section S1 [10]) yielded similar oscillations in the $S(Q, t)$ maps attributable to SAWs. Furthermore, a different pump wavelength ($\lambda_{opt}$ = 790 nm) produced very similar results (Supplemental Section S3 [10]). These findings suggest that the observed effect is universal, occurring in any sample that absorbs the pump light.

The surface phonons responsible for the observed DES patterns extend to values of $Q$ as large as 0.11 $nm^{-1}$, i.e. wavelengths as short as about 60 nm. This raises a question: how can a broadband spectrum of SAWs with nanoscale wavelengths be excited by the much longer wavelength pump?

A seemingly analogous effect was previously reported in diffuse scattering of hard x-rays following optical excitation of bulk phonons in GaAs containing ErAs nanoparticles [15,16]. In our experiment a similar effect could arise from the nanoscale inhomogeneity of the natural surface roughness. To test this hypothesis, we fabricated samples with variations of surface morphology. Atomic force microscopy reveals a smoother surface for the Pt layer grown on the Si substrate (rms roughness $\sigma_R$ = 0.6 nm) compared to the film grown on $Si_3N_4$ ($\sigma_R$ = 1.1 nm), see Supplemental Figs. S14 and S15 [10].

The differential DES images (Figs. 4(b) and 4(d)), measured at the same $F_{ex}$ (10 mJ/cm$^2$) and $t$ (350 ps), show a larger modulation amplitude for the sample with greater surface roughness. This trend is even more evident in Figure 4(c), where radial averages $S(Q, t = 350\ ps)$ are compared [17]. Similar results were obtained on multilayer Ta/Pt films grown on different substrates (Supplemental Section S4 [10]). These findings suggest that natural surface roughness plays a crucial role in enabling the optical excitation of high-wave-vector coherent SAWs. We identify two potential mechanisms. The first one is based on acoustic scattering from a rough surface [18]. When a laser pulse is absorbed in a sub-surface layer of a material with a perfectly flat surface, it launches a longitudinal acoustic pulse into the sample and another one propagating towards the surface, which gets immediately reflected from the surface with an opposite strain sign [19]. If the surface is rough, however, the reflection will be accompanied by scattering into both bulk and surface acoustic modes [18]. The second mechanism is based on optical scattering from surface roughness: the interference between the incident and scattered light leads to an inhomogeneous distribution of the absorbed intensity. This phenomenon has been studied in the context of laser-induced periodic surface structures forming when the fluence exceeds the material's damage threshold [20-23]. In our case, the inhomogeneous absorption may induce non-uniform thermal stresses, acting as a spatially random source of mechanical excitation and launching a broadband spectrum of SAWs. The first mechanism appears more likely, as we have not observed the polarization and wavelength dependencies characteristic of the second mechanism [21,22]. A direct test distinguishing the two mechanisms could be conducted on a sample consisting of an optical transparent layer (but opaque for the EUV probe) on an absorbing substrate. If the layer thickness is larger than the SAW wavelength, non-uniform absorption at a buried interface will not generate SAWs at the surface. In contrast, a longitudinal wave generated at the interface will propagate to the surface and get scattered into SAWs. Therefore, a fringe pattern in the DES signal is expected to appear after a time delay $\Delta t = \ell / v_L$, where $\ell$ is the thickness and $v_L$ the longitudinal sound velocity of transparent layer.

In either of the proposed mechanisms, a spatial Fourier component of surface roughness at an in-plane wave vector $\boldsymbol{Q}$ will generate counter-propagating SAWs. Without loss of generality, let us assume that $\boldsymbol{Q}$ is directed along the $x$-axis, and the corresponding Fourier component of the surface roughness is described by $u_0 cos(Qx + \theta)$. The SAW surface displacement can then be written as $u_{SAW} cos(Qvt + \varphi) cos(Qx + \theta)$, where $v(Q)$ is the SAW phase velocity. Since the excitation of SAWs is mediated by the surface roughness, their spatial phase is locked to that of the roughness itself. The scattering intensity at $Q$ is proportional to the square modulus of the Fourier-component of the surface displacement, i.e. $I_{DES}^{Pumped} \propto [u_0 + u_{SAW} cos(Qvt + \varphi)]^2$. Assuming the SAW displacement is much smaller than surface roughness, $u_{SAW} \ll u_0$, the time-dependent component of the scattering intensity becomes:

$$\Delta I_{DES} \propto 2 u_0 u_{SAW} cos(Qvt + \varphi), \qquad (2)$$

This expression describes both the circular fringe patterns in the $\boldsymbol{Q}$ space, with the period diminishing with increasing $t$ (Figs. 1(d)-1(f)), and the oscillations in $t$ at the SAW frequency $\omega = Qv$ (Figs. 2(a)-2(b)). The linear dependence of $\Delta I_{DES}$ on $F_{ex}$ (Fig. 2(d)) can also be explained by Eq.(2). Indeed, $u_{SAW}$ is due to thermal expansion, which scales linearly with the laser fluence $F_{ex}$ [24]. The physical interpretation of Eq.(2) is a phase-locked interference between the EUV field scattered by the surface roughness and the one scattered by the SAW. From the ratio of the DES fringes amplitude to the scattering intensity, we estimate that the ratio $u_{SAW}/u_0$ in our experiments ranges between 1 – 3%.

The phenomenon reported here enables access to surface phonon wave vectors beyond the reach of conventional optical scattering techniques. In this experiment, the accessible $Q$-range was limited by the experimental geometry and the size of the detector. Higher wave vectors can be reached either by acquiring diffuse scattering at larger angles relative to the specular reflection or by employing shorter values of $\lambda_{FEL}$ [25]. Although surface phonons with comparable wave vectors can also be generated using EUV TG [3-5], our method is significantly simpler and has the advantage of probing a wide $Q$-range simultaneously. Furthermore, the linear dependence of the signal on the EUV intensity suggests that the technique is compatible with table-top EUV sources. In our measurements, an EUV photon count in the range of $5\times10^{12}$ - $5\times10^{13}$ (Supplemental Table S1 [10]) provided a good signal-to-noise ratio. Since modern table-top high harmonic generation sources can deliver $10^{12}$-$10^{13}$ ph/s/eV [26], a DES image with similar counting contrast and probing bandwidth ($\Delta\lambda/\lambda \sim 10^{-3}$) could be acquired within minutes.

In summary, we have demonstrated that coherent SAWs with nanoscale wavelengths can be excited optically and probed using time-resolved DES, with both generation and detection facilitated by surface roughness [27]. This phenomenon is expected to occur universally in any material with nonzero surface roughness and absorption of the pump wavelength. Time-resolved DES thus provides a way to investigate surface phonons in a largely unexplored wave vector range. For example, there are no experimental data for surface phonon lifetime in this range, with the exception of a recent EUV TG study [5] reporting measurements for a single wave vector on a single material. Measuring the lifetime of high wave-vector surface phonons would offer insights into SAW attenuation mechanisms, such as phonon-phonon and electron-phonon interactions, as well as scattering processes involving surface roughness and grain boundaries, and would aid the development of SAW devices operating in the tens of GHz range [28,29]. Other applications may include characterizing elastic properties of ultra-thin films and subsurface layers [7]. For example, time-resolved DES would be well-suited for broad-band mechanical characterization of glass-forming materials available only as thin films [30]. Finally, since DES probes surface phonons over all azimuthal angles, this approach could be used to study high-frequency magnon-SAW coupling [31], where the hybridization of magnon and phonon modes depends on the angle between the SAW propagation direction and the magnetic field [32,33].

# Acknowledgements

The authors thank Plamen Krastev for providing the code for calculating surface phonon dispersion for a film stack containing a large number of layers. The contribution by A.A.M. and K.A.N. was supported by the Department of Energy, Office of Science, Office of Basic Energy Sciences, under award number DE-SC0019126. D.K. and C.G. acknowledge funding by the Deutsche Forschungsgemeinschaft (DFG) projects GU 535/9-1 and KS 62/3- 1. C.G. acknowledges funding from BMBF (05K24PSA) and DFG (NFDI 40/1).

# End Matter

*Data Analysis Procedure*

Four images were collected under different illumination conditions:

(a) signal image ($I_{FEL+Laser}$), collected at a fixed delay with both the FEL and pump laser illuminating the sample surface;
(b) laser background image ($I_{BG\ Laser}$), collected with only the optical laser active to account for optical stray radiation reaching the CCD detector;
(c) FEL background image ($I_{FEL}$), collected with only the FEL illumination;
(d) dark background image ($I_{BG}$ ), collected without any radiation inside the experimental chamber to measure the CCD detector's readout noise level.

For each signal image, background correction was performed as follows: $I_{\mathrm{DES}}^{Pumped} = I_{FEL+Laser} - I_{Laser}$; $I_{\mathrm{DES}}^{Static} = I_{FEL} - I_{BG}$. The resulting background-free images were then normalized by the total FEL intensity measured by the beamline photon transport intensity monitor, and the two normalized images were subtracted to obtain the differential scattered intensity image $\Delta I_{\mathrm{DES}}$, as described in the main text.

Finally, an isomorphic transformation was applied to correct for the stereographic projection of the diffraction pattern onto the flat CCD plane and the nonlinear coordinate warping caused by the angle between the specimen and the detector [34,35]. Figures 5(a) and 5(b) illustrate this correction for an image collected on the Pt/Al multilayer sample at $t$ = 250 ps: Figure 5(a) shows the DES data as collected on the CCD chip, while Figure 5(b) displays the corrected data in the $(Q_x, Q_y)$ plane. In the corrected image, the fringes become nearly perfectly circular. During a dynamical temporal scan, the images $I_{FEL}$ and $I_{BG}$ used for defining the static sample diffuse scattering were collected each 10 temporal points.

After applying the geometrical correction, the radial average shown in Figure 5(c) is obtained by integrating the scattered intensity over the angle at a given $Q = \sqrt{Q_x^2 + Q_y^2}$ from the center of the image (blue trace in Figure 5(c)). To account for low-frequency quasi-DC offset between different collected frames, which may arise from temporal drifts in CCD background noise level or residual laser stray light, at each time delay, a second-order polynomial background fit (dashed red line in Figure 5(c)) was calculated and subtracted from the radial average data. Figure 6 compares the azimuthal radial average map $S(Q,t)$ without (Fig. 6(a)) and with (Fig. 6(b)) low-frequency quasi-DC offset subtraction for the Pt/Al sample. One can see that the line-by-line low-frequency quasi-DC offset correction removes spurious intensity jumps likely caused by background noise fluctuations in the CCD detector, particularly in the ranges from 400 – 500 ps and 600 – 700 ps. On the other hand, the correction introduces minimal perturbation in the SAW dispersion, which is shown for both cases in Figures 6(c) and 6(d)**,** respectively.

It is important to note that while this procedure reduces background fluctuations in the experimental data, it may introduce artifacts near time zero (i.e., the moment when the optical laser and FEL pulses overlap) and during the first few tens of picoseconds of SAWs dynamic, when the fringe pattern in the detector plane is not fully developed. However, once the ring structure is well developed within the detector's field of view, the impact of these potential artifacts diminishes. Additionally, since the phonon dispersion curve is retrieved by applying a Fourier transform to the full dataset, any residual artifacts have a negligible effect on the final estimate of the surface phonon velocity.

*Wavevector and Frequency resolution*

The wavevector resolution $\Delta Q$ of our method is limited by the solid angle covered by the individual pixel of the detector. In the present case $\Delta Q = 2 \cdot (2 \cdot \pi / \lambda_{FEL}) \cdot \sin(\mathrm{atan}(d_{pxl}/D)/2)$ = $9.5 \times 10^{-5}$ nm$^{-1}$, where $d_{pxl}$ is the CCD pixels size (13.5 μm), $D$ is the sample to detector distance (50 mm) and $\lambda_{FEL}$ is the FEL radiation wavelength (17.8 nm). The final spectral resolution can also be affected by the FEL source bandwidth. At FERMI the relative bandwidth $\left(\frac{\Delta\lambda_{FEL}}{\lambda_{FEL}}\right)$ is of the order $10^{-3}$. Therefore, the point spread function due to the source bandwidth is negligible and it will affect the DES pattern only close to the detector edge [36]. The wavevector range is determined by the detector lateral dimension, its relative distance to the sample and its position relative to the reflected beam. The $Q$-range can be extended by moving the detector away from the specularly reflected beam.

The frequency resolution is limited by the range of the optical delay line, in the present case 1 ns (i.e. 1 GHz in the frequency domain). Higher resolution can be achieved by using a longer optical delay line. The frequency range is limited by the time duration of the pump and probe pulses, in the present case the use of sub-100 fs pulses in principle enables the detection of frequencies as large as ~10 THz, which exceeds the maximum extent of the surface phonon spectrum in most materials.

# Data availability

The data supporting the findings of this article are openly available [47], embargo periods may apply.

# Figures

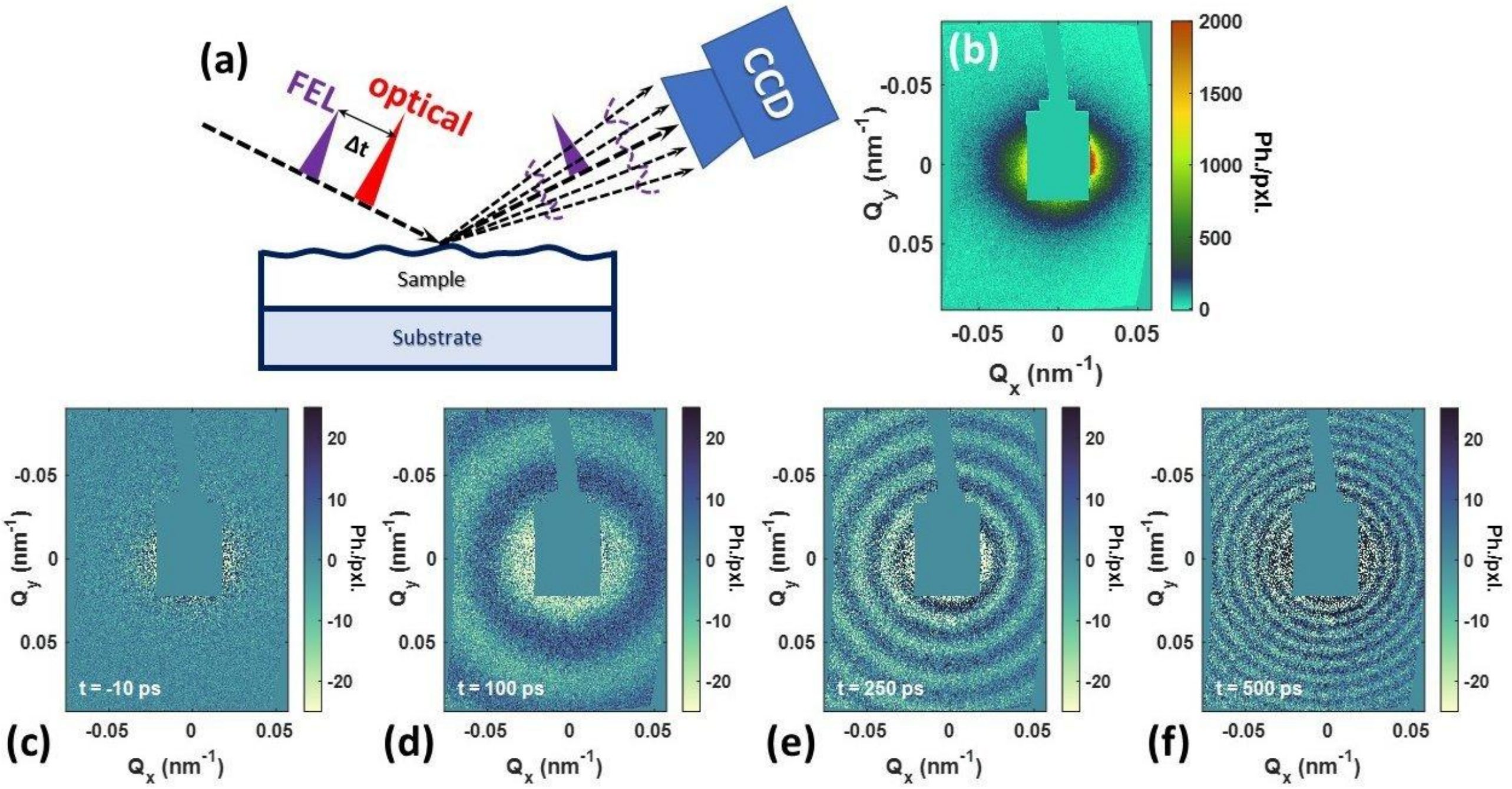

**Figure 1. (a)** Sketch of the experimental setup. **(b)** Static DES intensity pattern recorded by the CCD detector on the Pt/Al multilayer sample. The central rectangular region is the beam stop used to suppress the specular reflection. **(c) - (f)** $\Delta I_{\mathrm{DES}}\left(Q_x, Q_y, t\right)$ recorded at different pump-probe time delays.

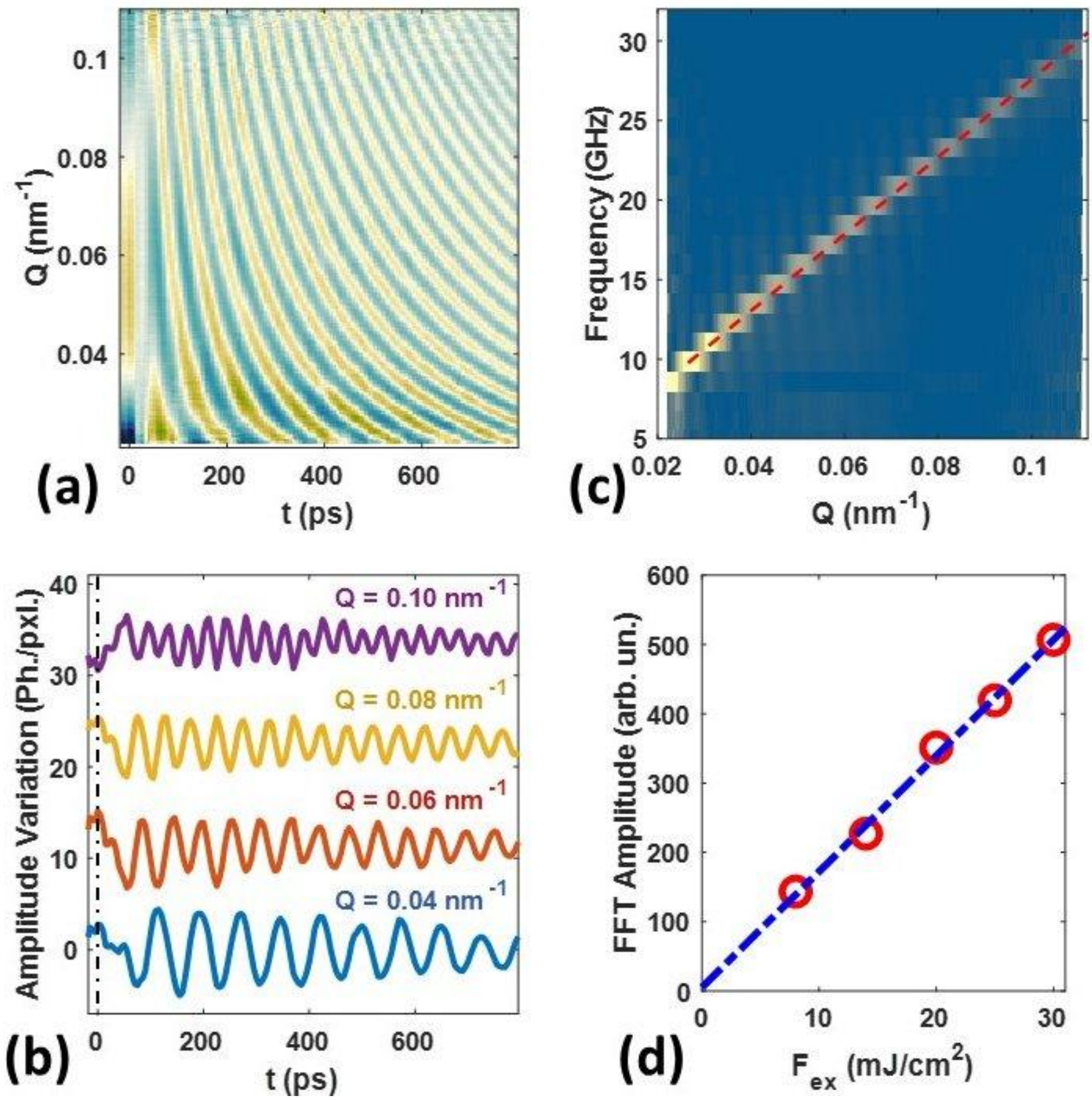

**Figure 2. (a)** Radial average of the DES intensity $S(Q,t)$ for the Pt/Al multilayer sample. **(b)** Time dependence of $S(Q,t)$ at selected values of $Q$ in the range 0.04 – 0.10 nm$^{-1}$ [11]. The vertical dash-dotted line denotes $t = 0$. **(c)** Temporal Fourier transform of the data in panel **(a)**. The red dashed line represents the calculated dispersion relation. **(d)** Dependence of the peak amplitude of the Fourier transform of $S(Q,t)$ on the laser fluence at $Q$ = 0.05 nm$^{-1}$.

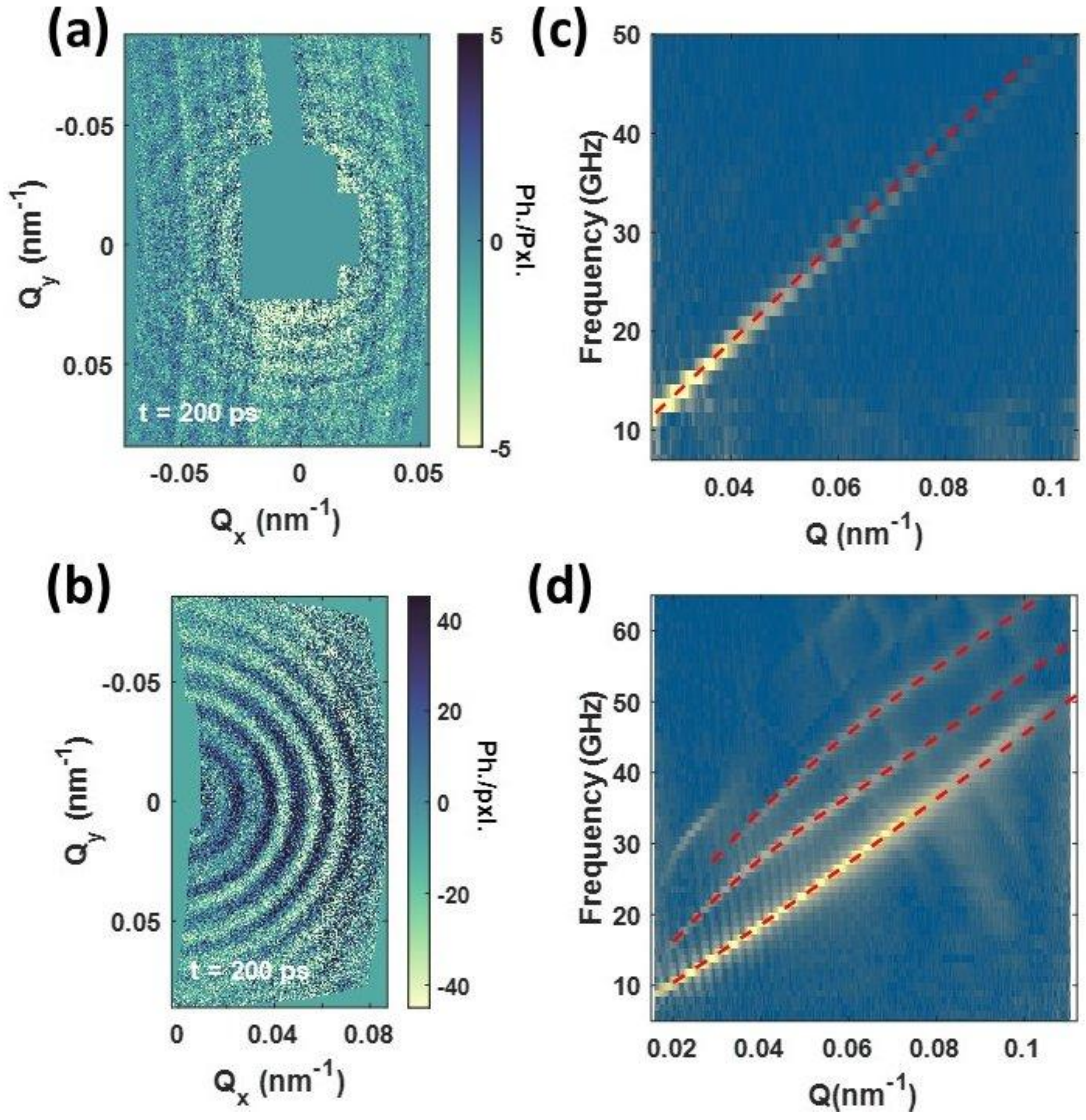

**Figure 3. (a)** $\Delta I_{\mathrm{DES}}\left(Q_x, Q_y, t = 200\ ps\right)$ for a (001) oriented GaAs crystal and **(b)** 100 nm thick Ti layer on a Si substrate. **(c)** Fourier transform of the $S(Q,t)$ map for the GaAs crystal, revealing a linear SAW dispersion (dashed line). **(d)** Fourier transform of $S(Q,t)$ for the Si/Ti sample. The dashed lines represent the calculated dispersion relations.

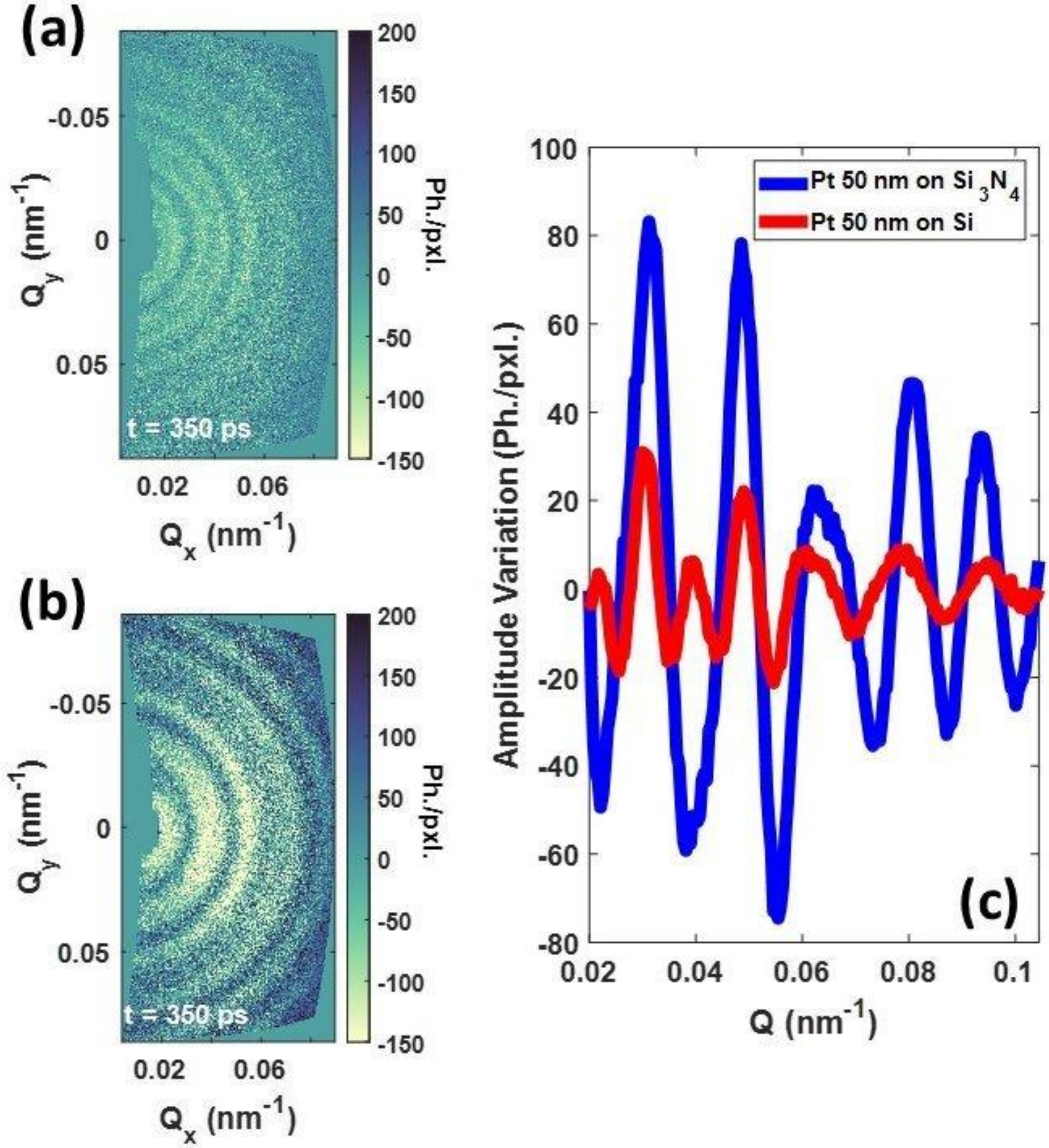

**Figure 4.** $\Delta I_{\mathrm{DES}}$ images for **(a)** Si/Pt sample and **(b)** Si/$Si_3N_4$/Pt sample recorded under identical conditions. **(c)** Radial averages $S(Q, t = 350\ ps)$ for the two samples.

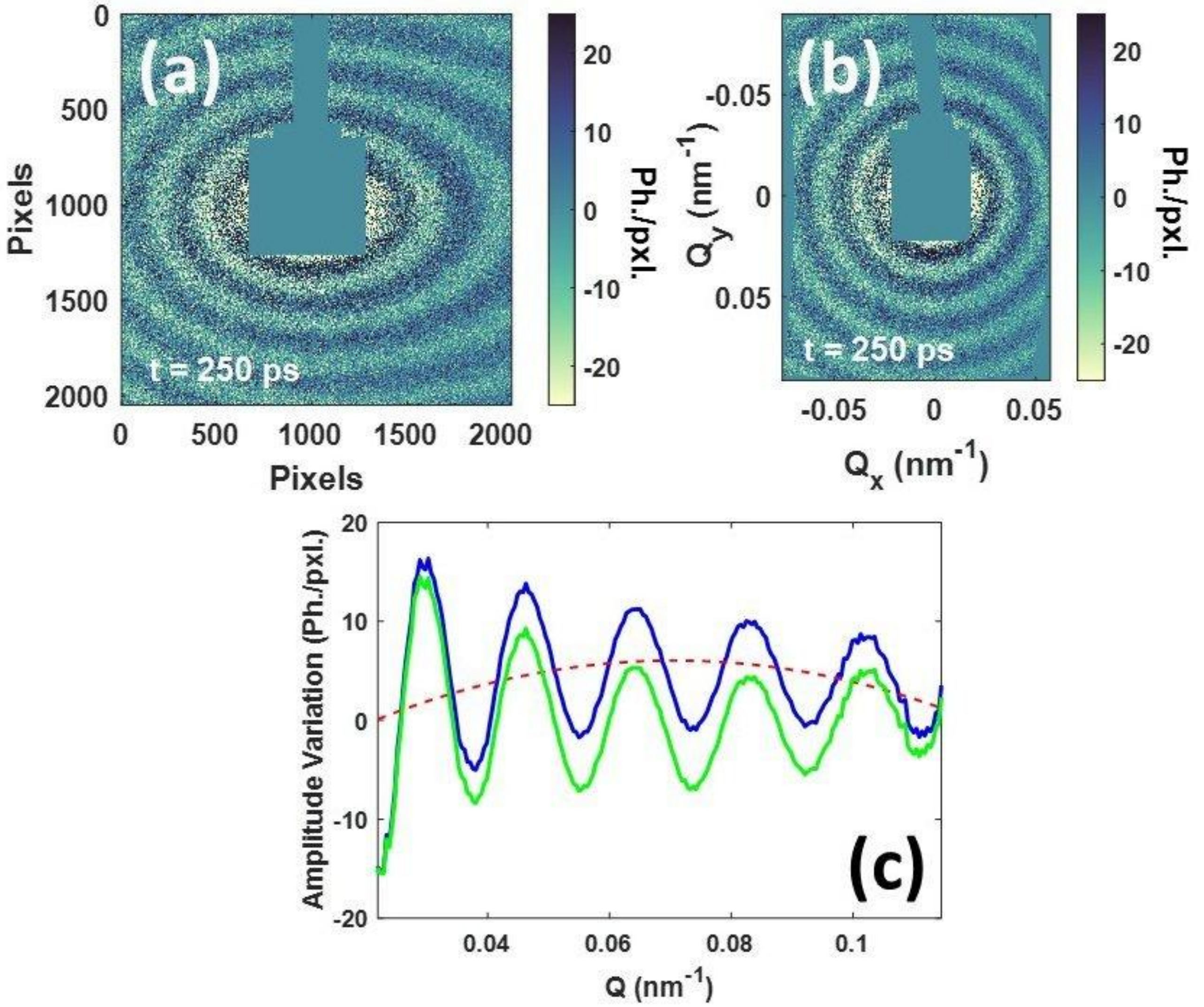

**Figure 5. (a)** $\Delta I_{\mathrm{DES}}$ image captured on the CCD chip **(b)** $\Delta I_{\mathrm{DES}}$ image after applying stereographic correction in the $(Q_x, Q_y)$ plane. **(c)** Azimuthal average of the data shown in panel **(b)** as a function of $Q$ (blue line). The dashed red line represents the low-frequency second-order polynomial background correction. The offset-compensated green trace for each time frame is used to construct the $S(Q, t)$ map.

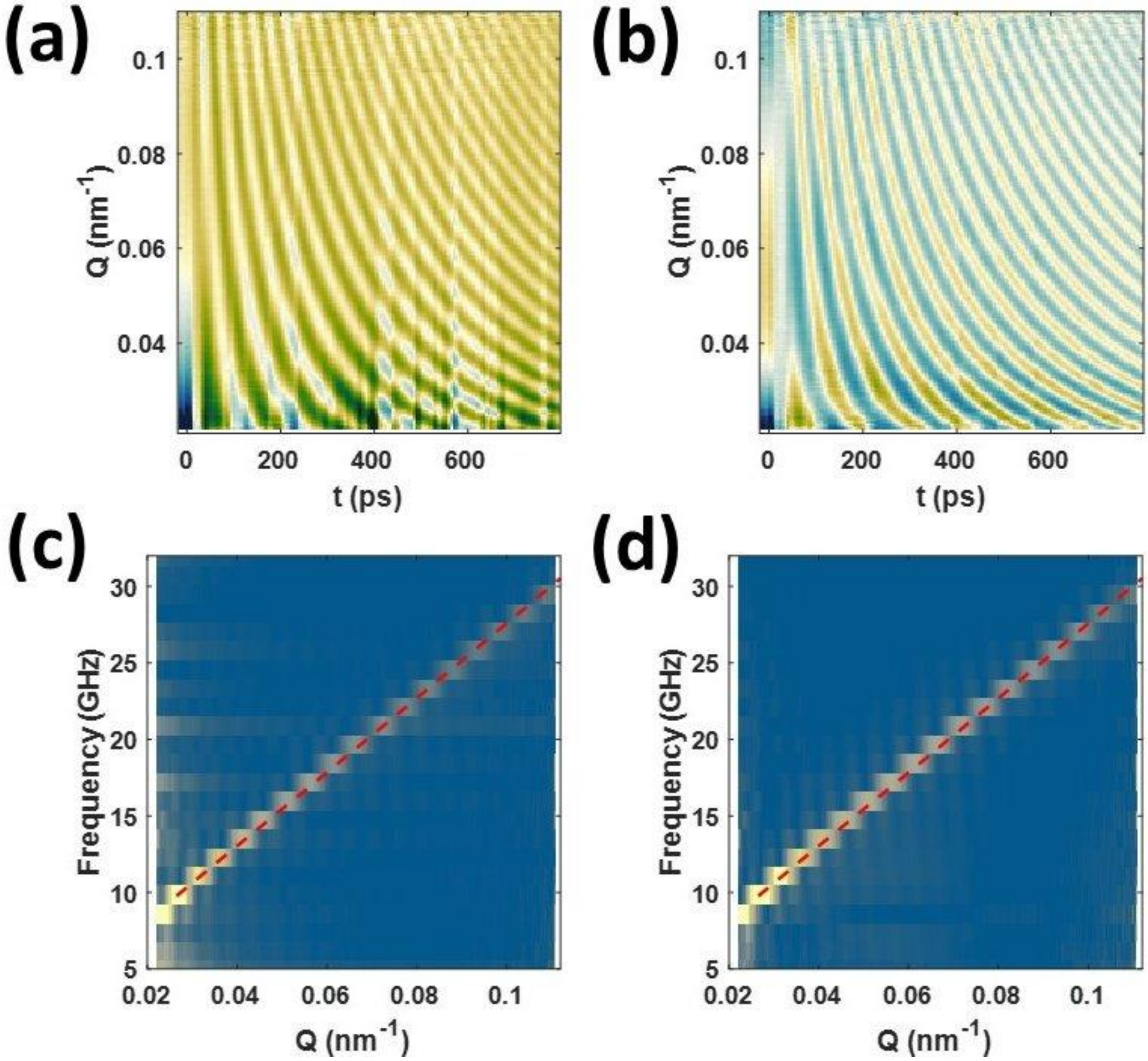

**Figure 6.** $S(Q,t)$ map for the Pt/Al multilayer sample without **(a)** - and with **(b)** - the line by line low frequency quasi-DC offset correction. **(c)** – **(d)** Fourier transform of the data in panel **(a)** and **(b)** respectively; the red dashed line represents the calculated dispersion relation.

# Time-domain extreme ultraviolet diffuse scattering spectroscopy of nanoscale surface phonons
# - *Supplemental Material* -

F. Capotondi[1,*,†], A.A. Maznev[2,*,‡], F. Bencivenga[1,*], S. Bonetti[3], D. Fainozzi[1,4], D. Fausti[1,5,6], L. Foglia[1], C. Gutt[7], N. Jaouen[8,3], D. Ksenzov[7], C. Masciovecchio[1], Keith A. Nelson[2], I. Nikolov[1], M. Pancaldi[1,3], E. Pedersoli[1], B. Pfau[9], L. Raimondi[1,10], F. Romanelli[11], R. Totani[1], M. Trigo[12,13]

[1]*Elettra Sincrotrone Trieste, Strada Statale 14, km 163.5, 34149 Basovizza, TS, Italy.*
[2]*Department of Chemistry, Massachusetts Institute of Technology, Cambridge, Massachusetts 02139, USA.*
[3] *Department of Molecular Sciences and Nanosystems, Ca' Foscari University of Venice, Venice, Italy*
[4] *Institute of Applied Physics, University of Bern, Sidlerstrasse 5, CH-3012 Bern, Switzerland*
[5]*Department of Physics, Università degli Studi di Trieste, 34127 Trieste, Italy*
[6]*Department of Physics, University of Erlangen-Nürnberg, 91058 Erlangen, Germany*
[7]*Department of Physics, University of Siegen, Walter-Flex-Strasse 3, 57072 Siegen, Germany.*
[8] *Synchrotron SOLEIL, Saint-Aubin, Boite Postale 48, 91192, Gif-sur-Yvette Cedex, France.*
[9] *Max Born Institute, Max-Born-Straße 2A, 12489 Berlin, Germany.*
[10] *Advanced Light Source, Lawrence Berkeley National Lab, Berkeley, CA 94720, USA*
[11] *Department of Mathematics, Informatics and Geosciences, Università degli Studi di Trieste, 34127 Trieste, Italy*
[12] *Stanford Institute for Materials and Energy Sciences, SLAC National Accelerator Laboratory, Menlo Park, California 94025, USA*
[13] *Stanford PULSE Institute, SLAC National Accelerator Laboratory, Menlo Park, California 94025, USA*

** These authors contributed equally to this work.*
*† Contact author: flavio.capotondi@elettra.eu*
*‡ Contact author: alexei.maznev@gmail.com*

## S1. EUV diffuse scattering on additional samples

This section presents additional $S(Q,t)$ maps and the corresponding dispersion curves for various samples measured during the experimental campaign as well as surface roughness measurements obtained via atomic force microscopy (AFM). Table S1 details the experimental conditions for each sample, with all experiments conducted using a 395 nm optical laser pulse for excitation and probing the dynamics with free electron laser (FEL) extreme ultraviolet (EUV) radiation at 17.8 nm.

| Sample | Laser energy density (mJ/cm²) | FEL energy density per shot (mJ/cm²) | Total Number of input photons per acquisition | Total Number of recorded photons on CCD detectors | Surface scattering efficiency | $\sigma_R$ (nm) | $\Lambda$ (nm) |
|---|---|---|---|---|---|---|---|
| Si/Co (75 nm)/Ta(2 nm) | 12.6±0.3 | 0.83±0.02 | 2.1 x10$^{13}$ | 2.25±0.05 x10$^{9}$ | 1.07±0.02 x10$^{-4}$ | 0.7 | 11±1 |
| Si/Ta (75 nm) | 23.1±0.3 | 1.45±0.02 | 1.9 x10$^{13}$ | 1.38±0.05 x10$^{9}$ | 7.2±0.3 x10$^{-5}$ | 0.8 | 9±1/6±1 |
| Si/Cr(7nm)/Au(75 nm) | 12.6±0.3 | 0.83±0.02 | 5.3 x10$^{12}$ | 7.6±0.1 x10$^{9}$ | 4.0±0.1 x10$^{-3}$ | 2.50 | 33±5 |
| Si | 52.8±0.7 | 0.90±0.02 | 3.5x10$^{13}$ | 5.7±0.3 x10$^{7}$ | 1.63±0.05 x10$^{-6}$ | 0.2 | 18±2 |
| Ge | 6.75±0.05 | 0.72±0.01 | 4.5x10$^{13}$ | 6.6±0.2 x10$^{8}$ | 1.46±0.04 x10$^{-5}$ | 0.35 | 22±2 |

**Table S1.** Experimental conditions used during the time resolved DES experiments and surface roughness of various samples as inferred from AFM image analysis.

Figures S1, S3, S5, S7, S9 present the AFM images of the five additional samples. The RMS surface roughness $\sigma_R$ listed in Table S1 ranges from 0.2 nm (bulk Si) to 2.5 nm (Au thin film), while the correlation length $\Lambda$, defined as the distance at which the autocorrelation function decays to 1/e of its initial value [38], is in the 10 – 30 nm range. Time-dependent $S(Q,t)$ intensity maps and the corresponding Fourier transform maps showing SAW phonon dispersion are reported in Figures S2, S4, S6, S8, S10.

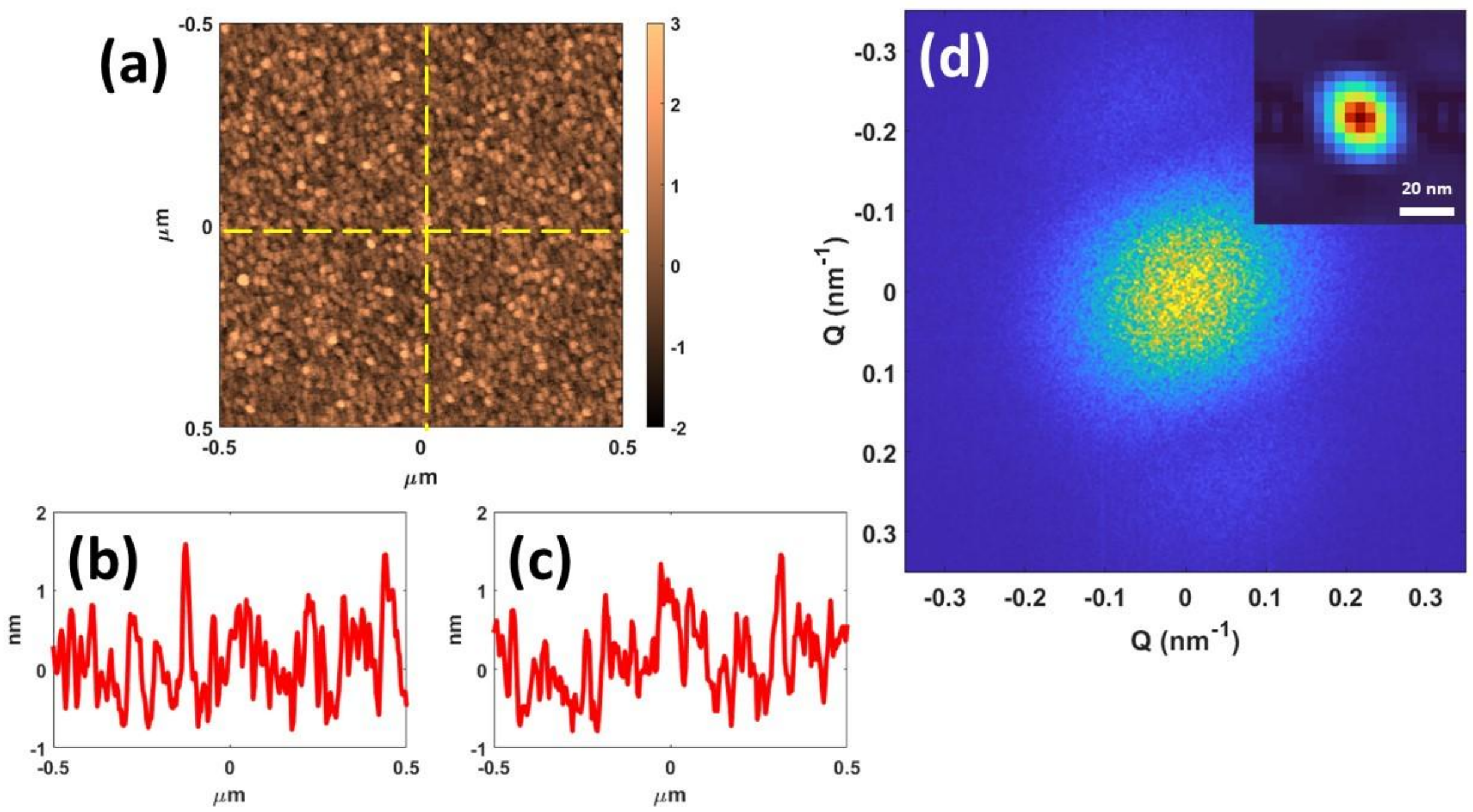

**Figure S1. (a)** 1x1 μm² AFM image of Si/Co (75 nm)/Ta(2 nm) sample, where the 2 nm Ta top layer was deposited to prevent Co oxidation. **(b)** Horizontal and **(c)** vertical profiles along the dashed yellow lines in panel **(a)**. Statistical analysis of these height profiles yields a $\sigma_R$ of 0.7 nm. **(d)** The Fourier transform of the AFM image. The top-right inset displays the central part of the autocorrelation function, revealing a nearly isotropic roughness correlation length of 11±1 nm.

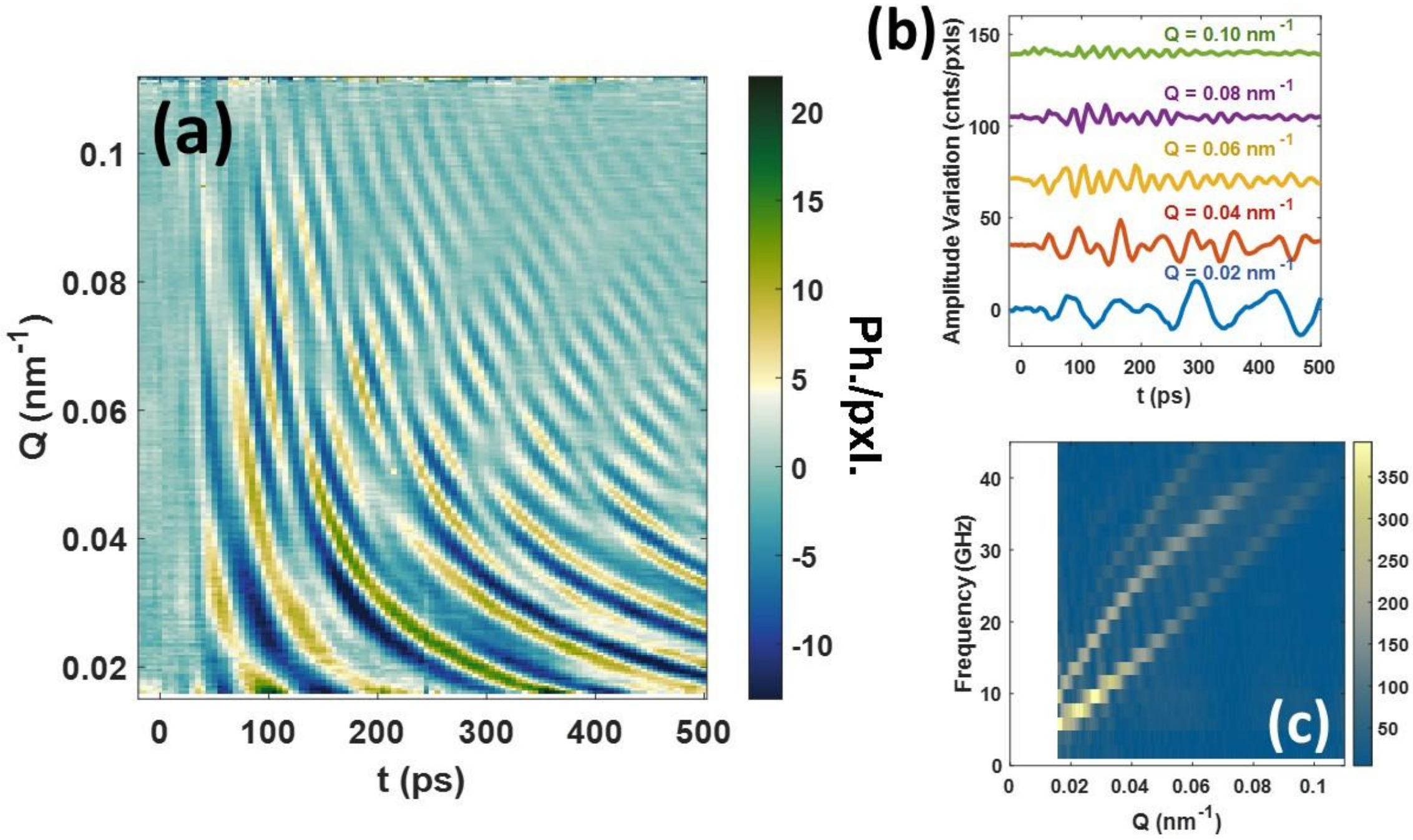

**Figure S2. (a)** Radially averaged differential diffuse scattering intensity $S(Q,t)$ as a function of $Q$ and $t$ for the Si/Co (75 nm)/Ta(2 nm) sample. **(b)** Time dependence of $S(Q,t)$ at selected values of $Q$ in the range 0.02 – 0.10 nm$^{-1}$. **(c)** Fourier transform of the data in panel **(a)**.

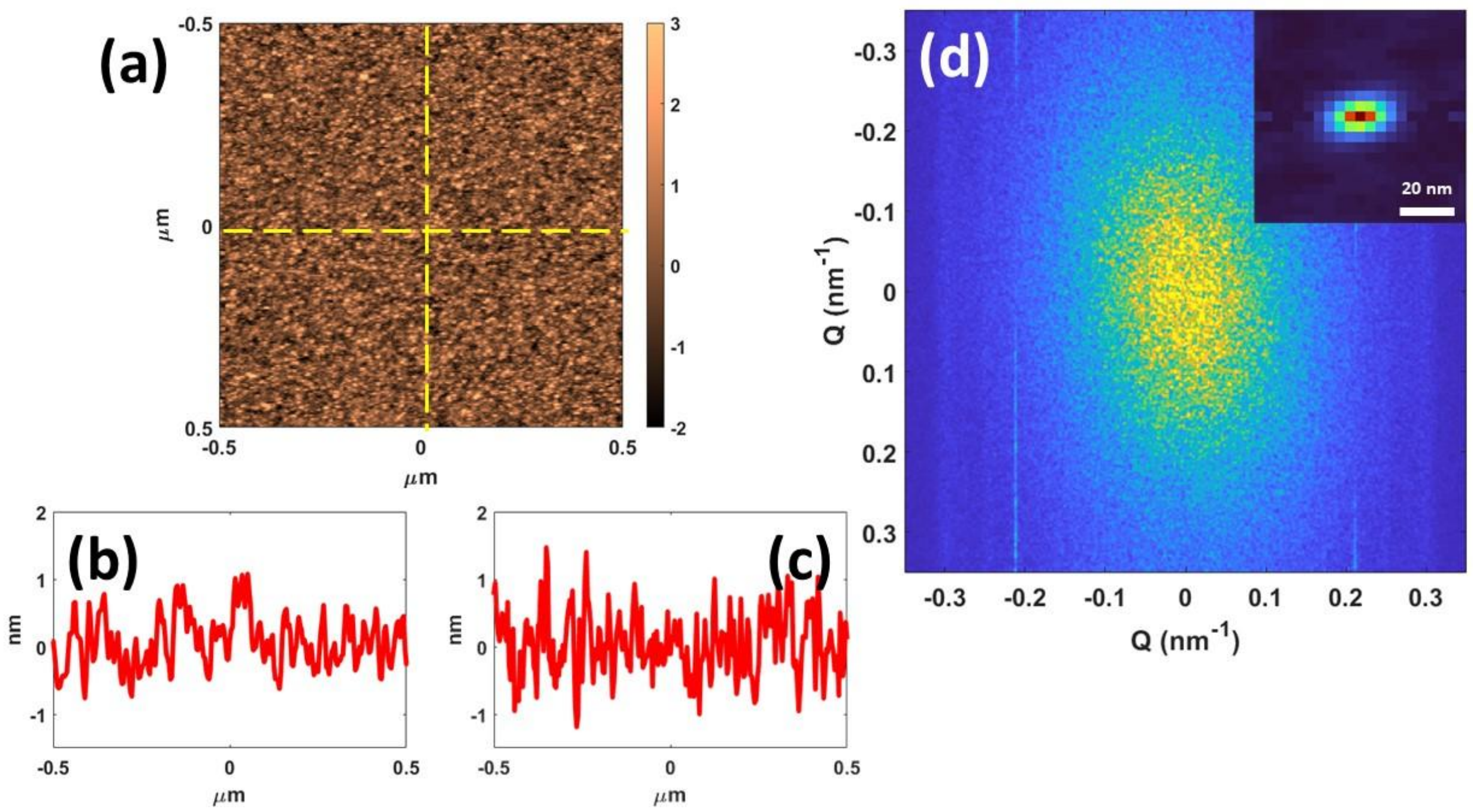

**Figure S3. (a)** 1x1 μm$^2$ AFM image of Si/Ta (75 nm) sample. **(b)** Horizontal and **(c)** vertical profiles along the dashed yellow lines in panel **(a)**. Statistical analysis of these height profiles yields a $\sigma_R$ of 0.8 nm. **(d)** The Fourier transform of the AFM image. The top-right inset provides the central part of the autocorrelation function, revealing an anisotropic correlation length of 9±1 nm in the horizontal direction and 6±1 nm in the vertical direction.

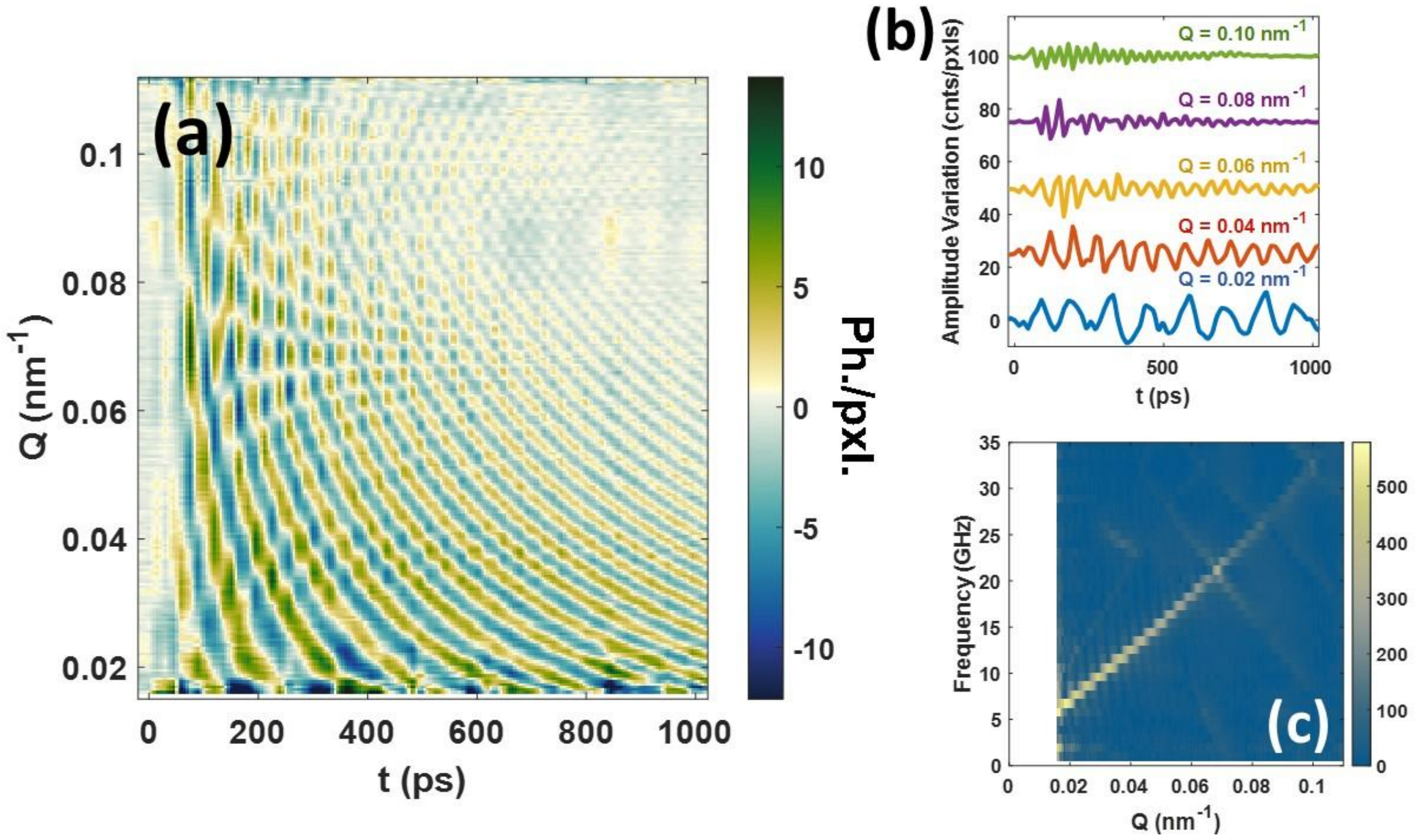

**Figure S4. (a)** Radially averaged differential diffuse scattering intensity $S(Q,t)$ as a function of $Q$ and $t$ for the Si/Ta (75 nm) sample. **(b)** Time dependence traces at selected values of $Q$ in the range 0.02 – 0.10 nm$^{-1}$. **(c)** Fourier transform of the data in panel **(a)**.

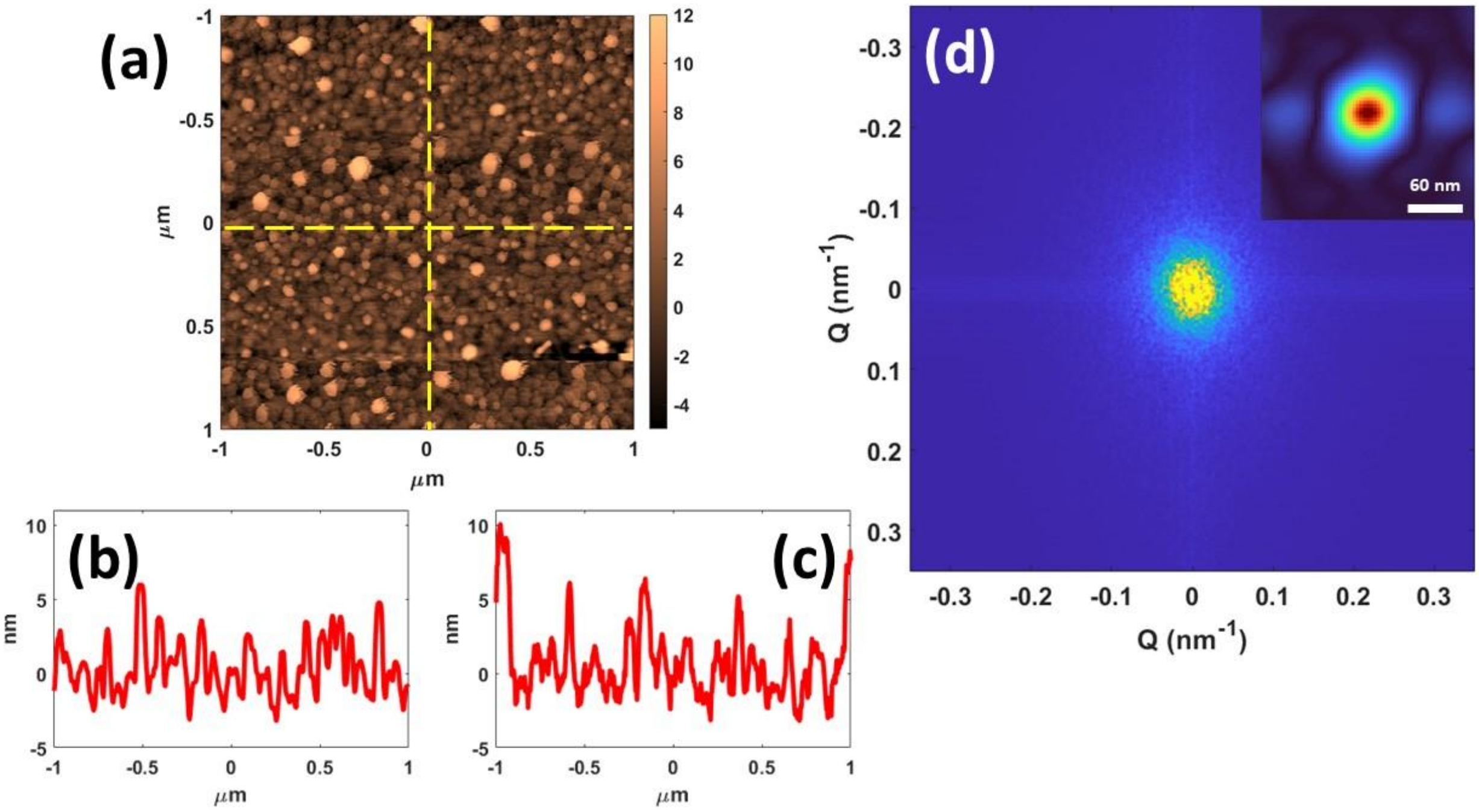

**Figure S5. (a)** 2x2 μm$^2$ AFM image of Si/Cr(7nm)/Au(75 nm) sample. The 7 nm Cr interlayer was deposited as an adhesion layer to facilitate the sputtering deposition of the Au film on the Si substrate. **(b)** Horizontal and **(c)** vertical profiles taken along the dashed yellow lines in panel **(a)**. Statistical analysis of these height profiles yields a $\sigma_R$ of 2.5 nm. **(d)** The Fourier transform of the AFM image. The top-right inset displays the central part of the autocorrelation function, revealing a nearly isotropic roughness correlation length of 33±5 nm.

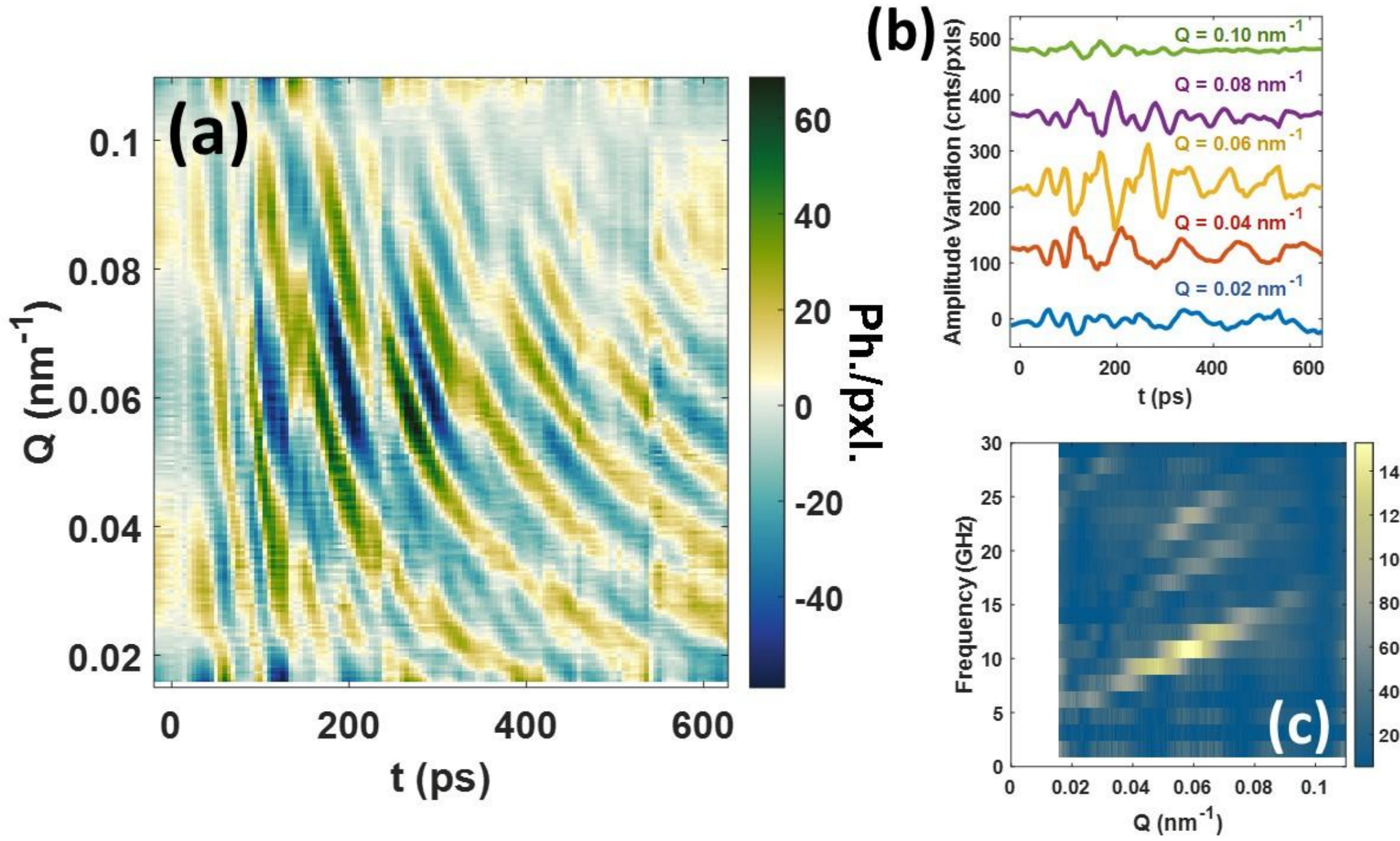

**Figure S6. (a)** Radially averaged differential diffuse scattering intensity $S(Q,t)$ as a function of $Q$ and $t$ for the Si/Cr(7nm)/Au(75 nm) sample. **(b)** Time dependence traces at selected values of $Q$ in the range 0.02 – 0.10 nm$^{-1}$. **(c)** Fourier transform of the data in panel **(a)**.

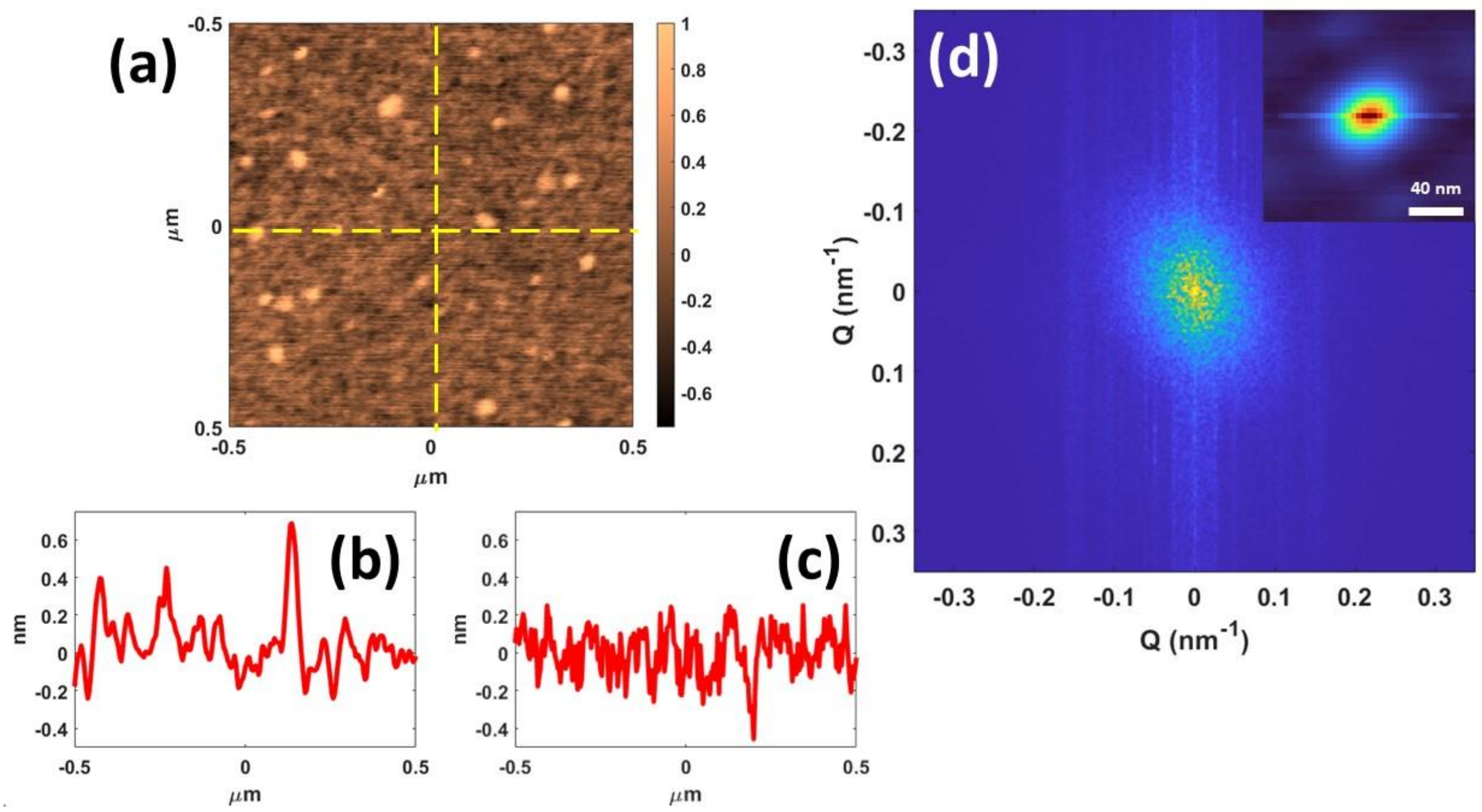

**Figure S7.** **(a)** 1x1 μm$^2$ AFM image of Si sample. **(b)** Horizontal and **(c)** vertical profiles taken along the dashed yellow lines in panel **(a)**. Statistical analysis of these height profiles yields a $\sigma_R$ of 0.2 nm. **(d)** The Fourier transform of the AFM image. The top-right inset displays the central part of the autocorrelation function, revealing a nearly isotropic roughness correlation length of 18±2 nm.

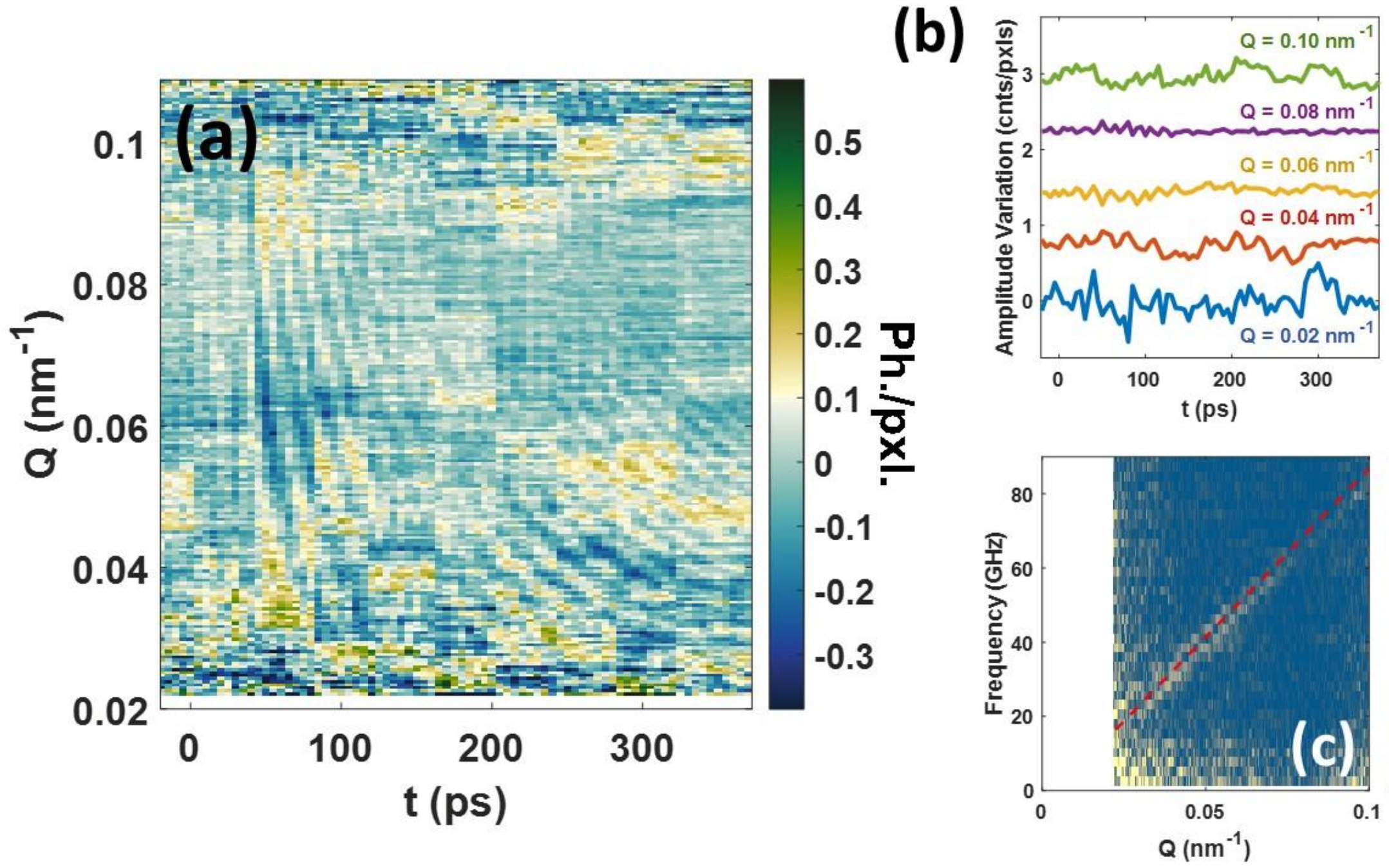

**Figure S8.** **(a)** Radially averaged differential diffuse scattering intensity $S(Q,t)$ as a function of $Q$ and $t$ for the Si bulk sample. **(b)** Time dependence traces at selected values of $Q$ in the range 0.02 – 0.10 nm$^{-1}$. **(c)** Fourier transform of the data in panel **(a),** with the dashed red line indicating the linear dispersion of the first Rayleigh mode, corresponding to a surface phonon propagation velocity of 5100±50 m/s. The measured value is in good agreement with the average Rayleigh wave velocity for bulk Si, 5000 m/s, calculated by averaging the SAW velocities for the [100] and [110] wavevector directions based on ref. [39].

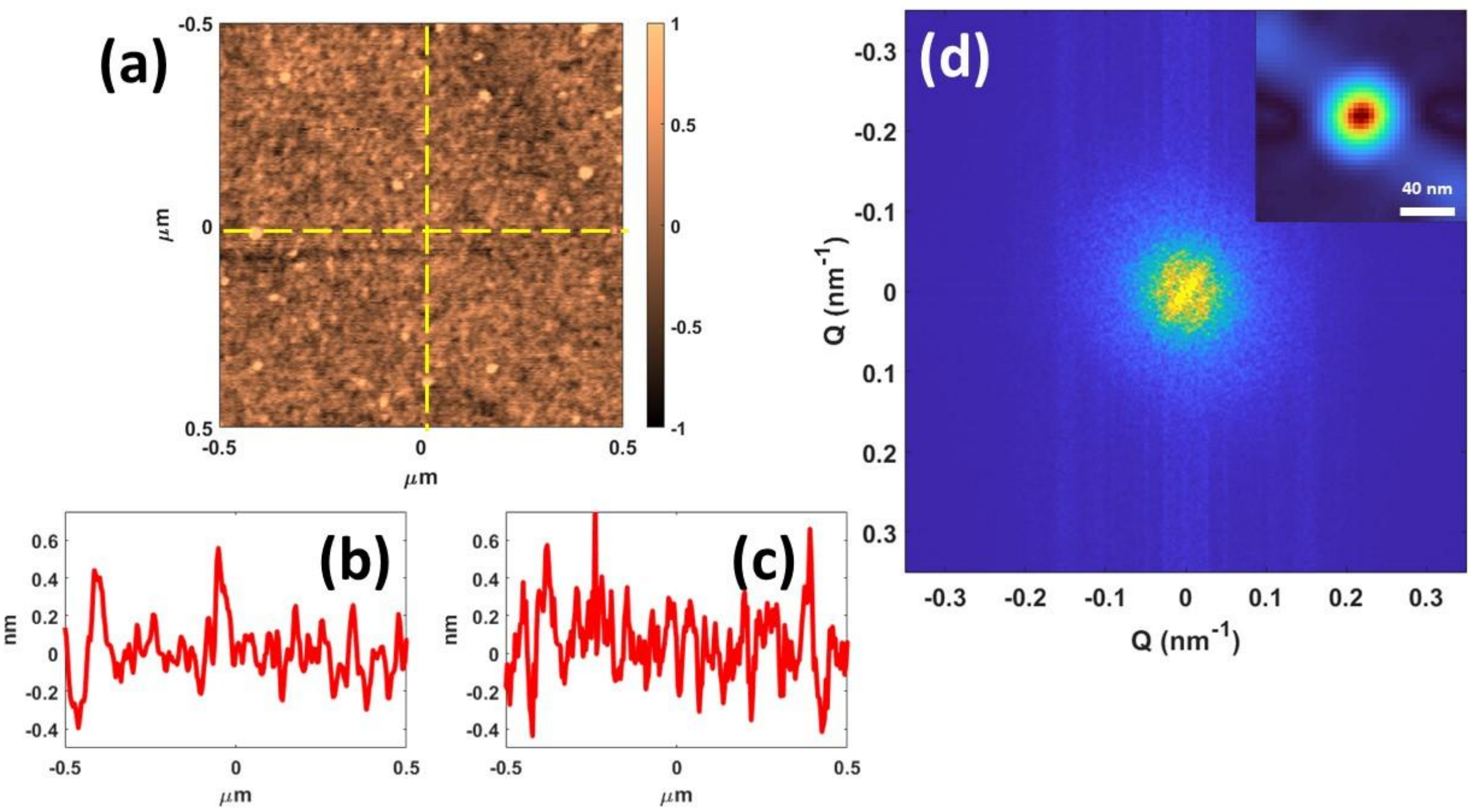

**Figure S9. (a)** 1x1 $\mu m^2$ AFM image of Ge sample. **(b)** Horizontal and **(c)** vertical profiles taken along the dashed yellow lines in panel **(a)**. Statistical analysis of these height profiles yields a $\sigma_R$ of 0.35 nm. **(d)** The Fourier transform of the AFM image. The top-right inset displays the central part of the autocorrelation function, revealing a nearly isotropic roughness correlation length of 20±2 nm.

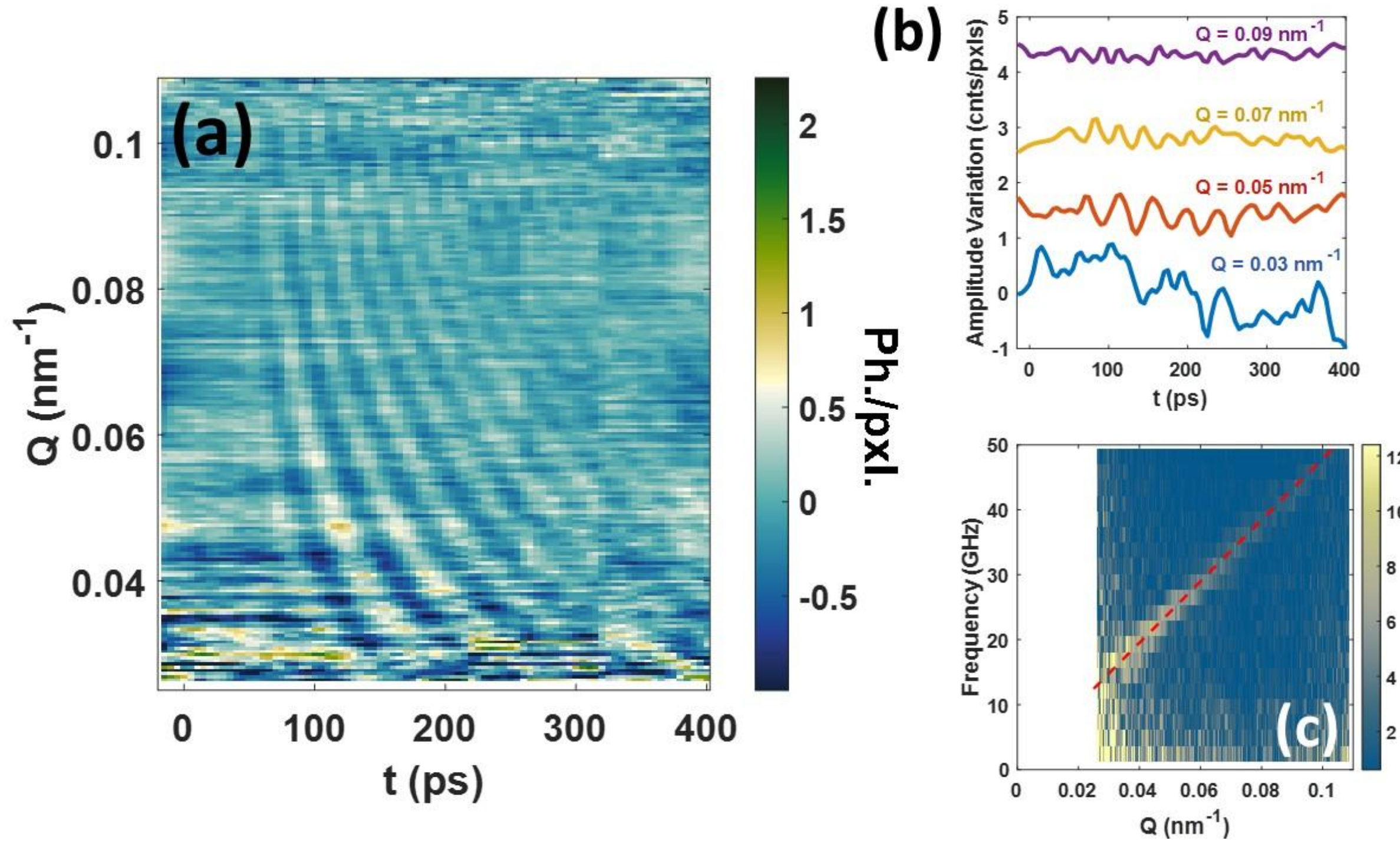

**Figure S10. (a)** Radially averaged differential diffuse scattering intensity $S(Q,t)$ as a function of $Q$ and $t$ for the Ge bulk sample. **(b)** Time dependence traces at selected values of $Q$ in the range 0.03 – 0.09 $nm^{-1}$. **(c)** Fourier transform of the data in panel **(a)**, with the dashed red line indicating the linear dispersion of the first Rayleigh mode, corresponding to a surface phonon propagation velocity of 3100±50 m/s. The measured value is in agreement with the average Rayleigh wave velocity for bulk Ge, 3000 m/s, calculated by averaging the SAW velocities for the [100] and [110] wavevector directions based on ref [39].

For the sake of completeness, Figures S11 - S15 report the AFM morphological characterizations of the samples presented in the main text. Figures S16 - S19 present time-dependent $S(Q,t)$ intensity maps and the corresponding Fourier transform maps showing SAWs phonon dispersion for the GaAs, Si/Ti(100nm), Si/Pt(50nm), Si/Si3N4(100nm)/Pt(100nm) samples, respectively. Note the prominence of the second SAW mode in time Fourier transform of the maps $S(Q,t)$ for Pt film grown on Si substrate (Fig. S18(c)) with respect to the one grown on Si/Si3N4(100nm) substrate (Fig. S19(c)).
All the thin film samples reported in Table S1 as well Si/Ti (100nm) and Si/Pt(50nm), Si/Si3N4(100nm)/Pt(100nm) samples were produced with electron-beam evaporation, while the multilayer sample was prepared with magnetron sputtering.

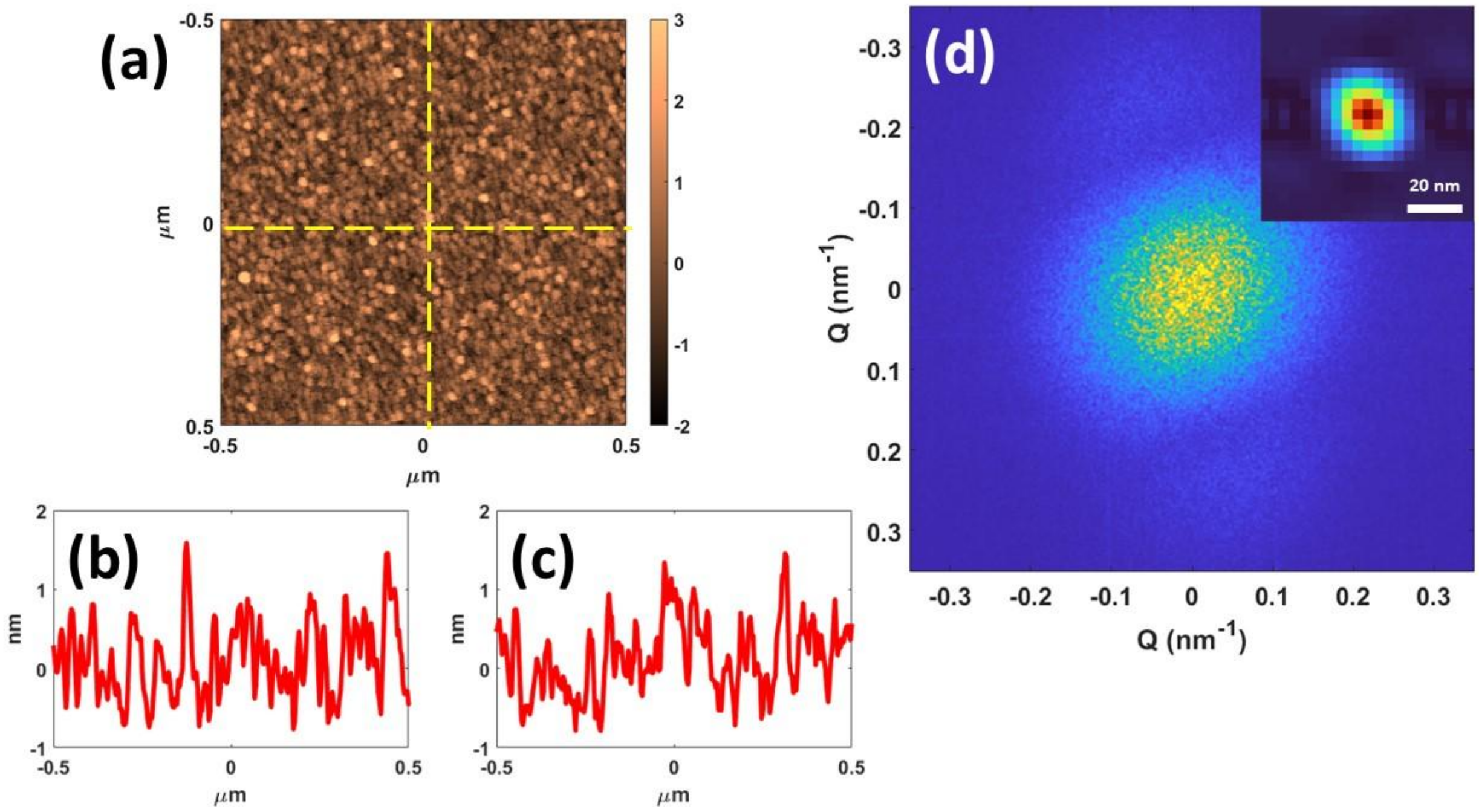

**Figure S11. (a)** 1x1 μm$^2$ AFM image of [Pt(4nm)/Al(4nm)]×4 multilayer stack grown on a Si/$SiO_2$ substrate. **(b)** Horizontal and **(c)** vertical profiles taken along the dashed yellow lines in panel **(a)**. Statistical analysis of these height profiles yields a $\sigma_R$ of 0.6 nm. **(d)** The Fourier transform of the AFM image. The top-right inset displays the central part of the autocorrelation function, revealing a nearly isotropic roughness correlation length of 14±2 nm.

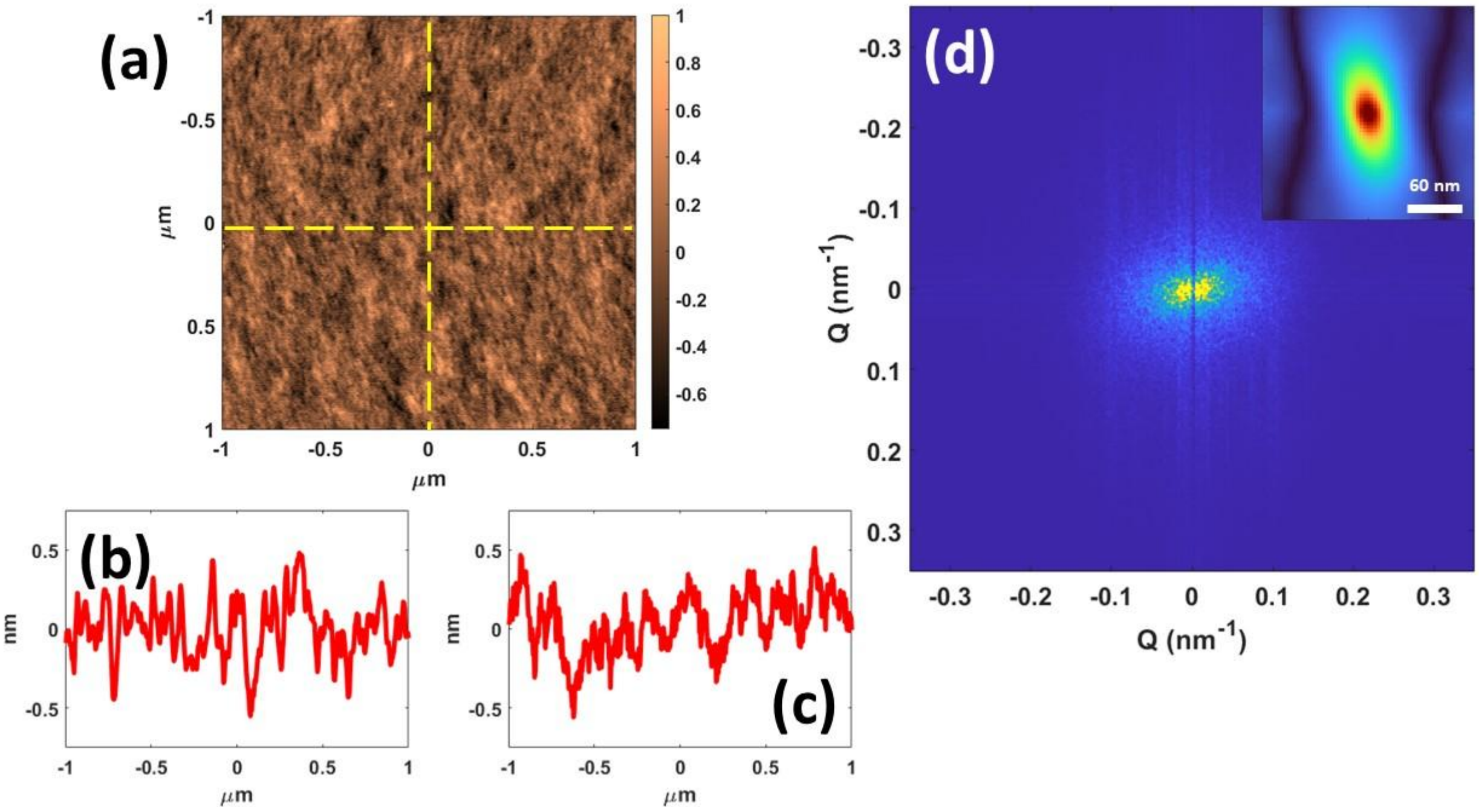

**Figure S12. (a)** 1x1 µm$^2$ AFM image of (001) oriented GaAs crystal. **(b)** Horizontal and **(c)** vertical profiles taken along the dashed yellow lines in panel **(a)**. Statistical analysis of these height profiles yields a $\sigma_R$ of 0.2 nm. **(d)** The Fourier transform of the AFM image. The top-right inset displays the central part of the autocorrelation function, revealing an anisotropic correlation length of 19±2 nm in the horizontal direction and 42±5 nm in the vertical direction.

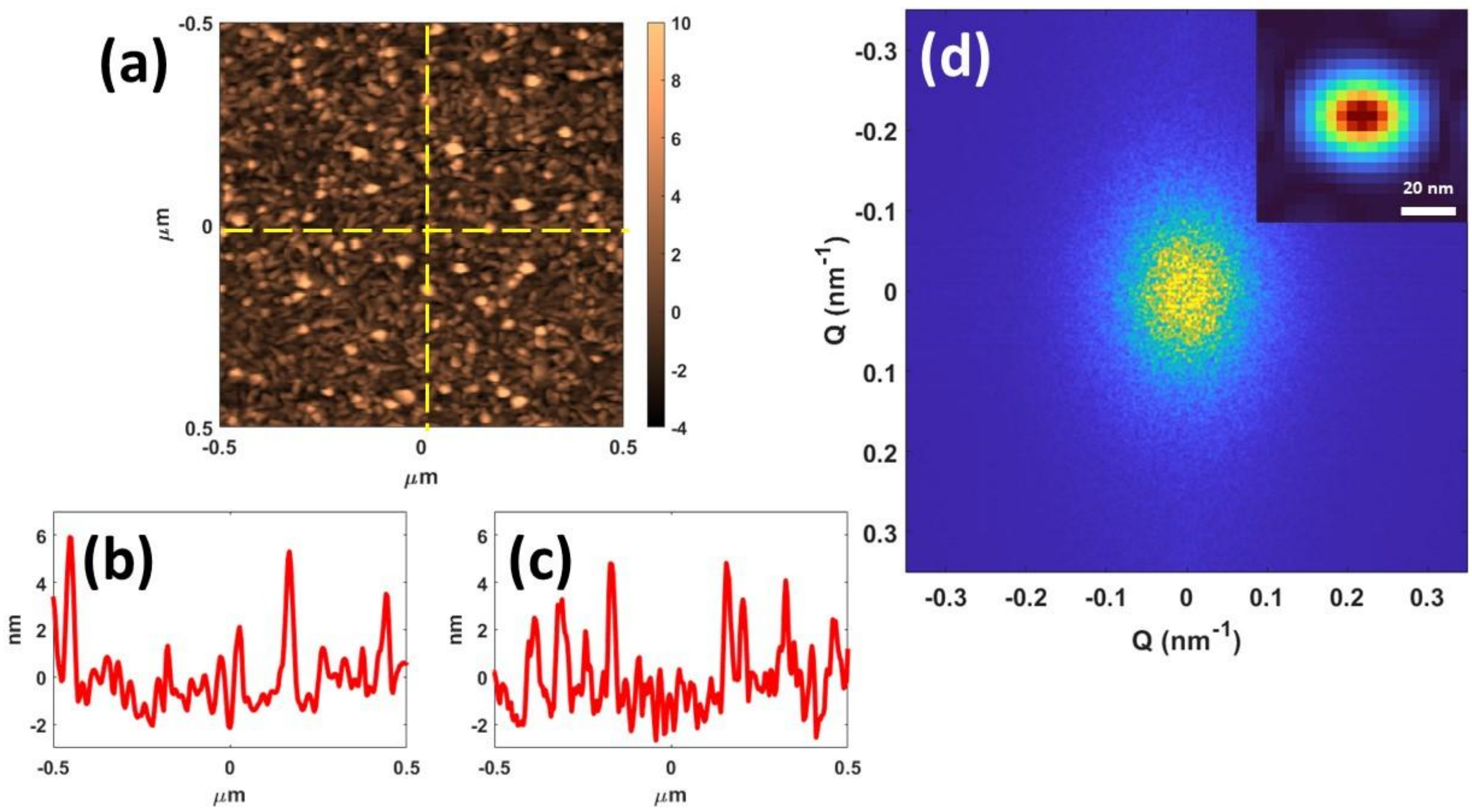

**Figure S13. (a)** 1x1 µm$^2$ AFM image of Si/Ti (100 nm) sample. **(b)** Horizontal and **(c)** vertical profiles taken along the dashed yellow lines in panel **(a)**. Statistical analysis of these height profiles yields a $\sigma_R$ of 2.0 nm. **(d)** The Fourier transform of the AFM image. The top-right inset displays the central part of the autocorrelation function, revealing a nearly isotropic roughness correlation length of 15±2 nm.

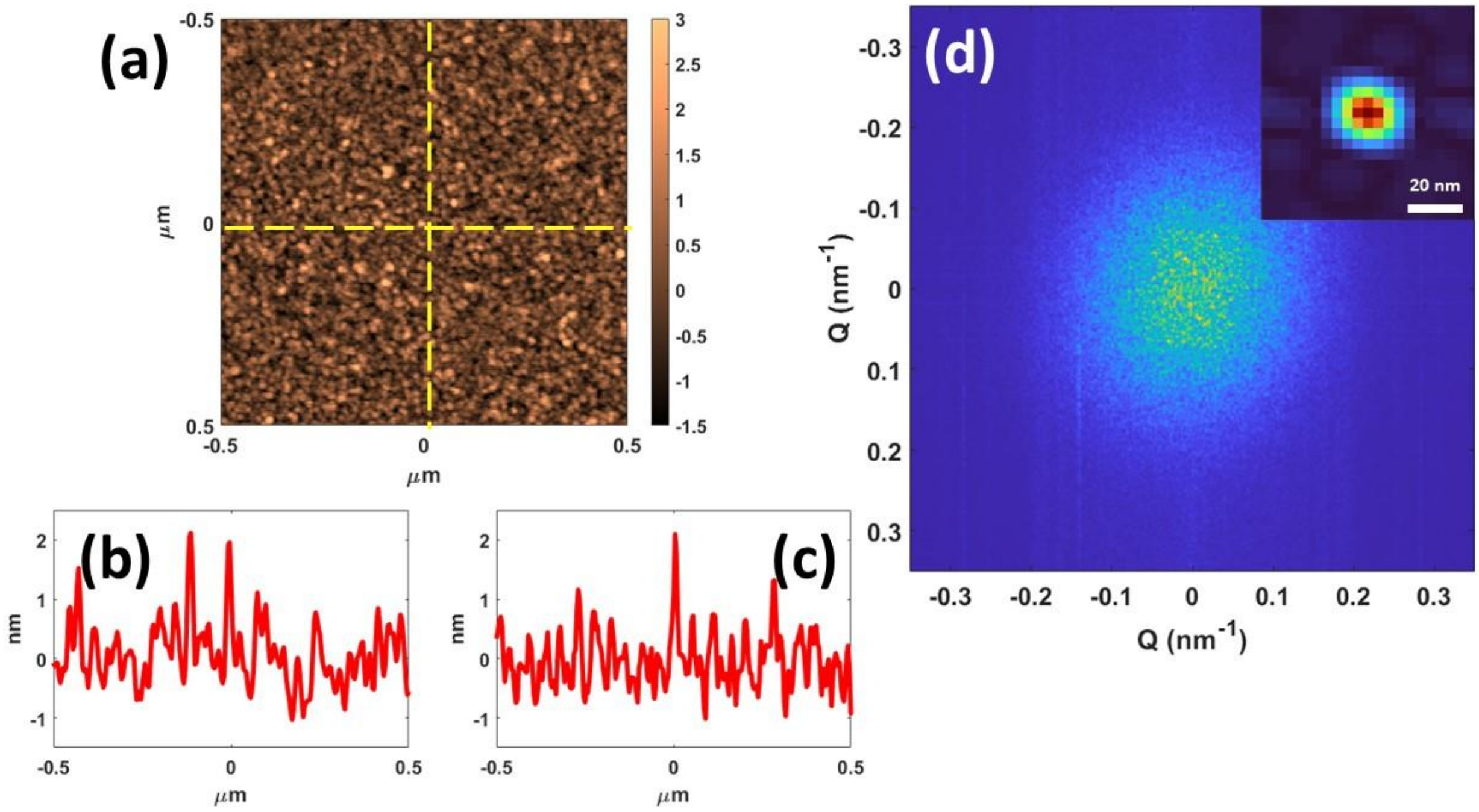

**Figure S14. (a)** 1x1 µm$^2$ AFM image of Si/Pt(50 nm) sample. **(b)** Horizontal and **(c)** vertical profiles taken along the dashed yellow lines in panel **(a)**. Statistical analysis of these height profiles yields a $\sigma_R$ of 0.6 nm. **(d)** shows the Fourier transform of the AFM image. The top-right inset displays the central part of the autocorrelation function, revealing a nearly isotropic roughness correlation length of 10±1 nm.

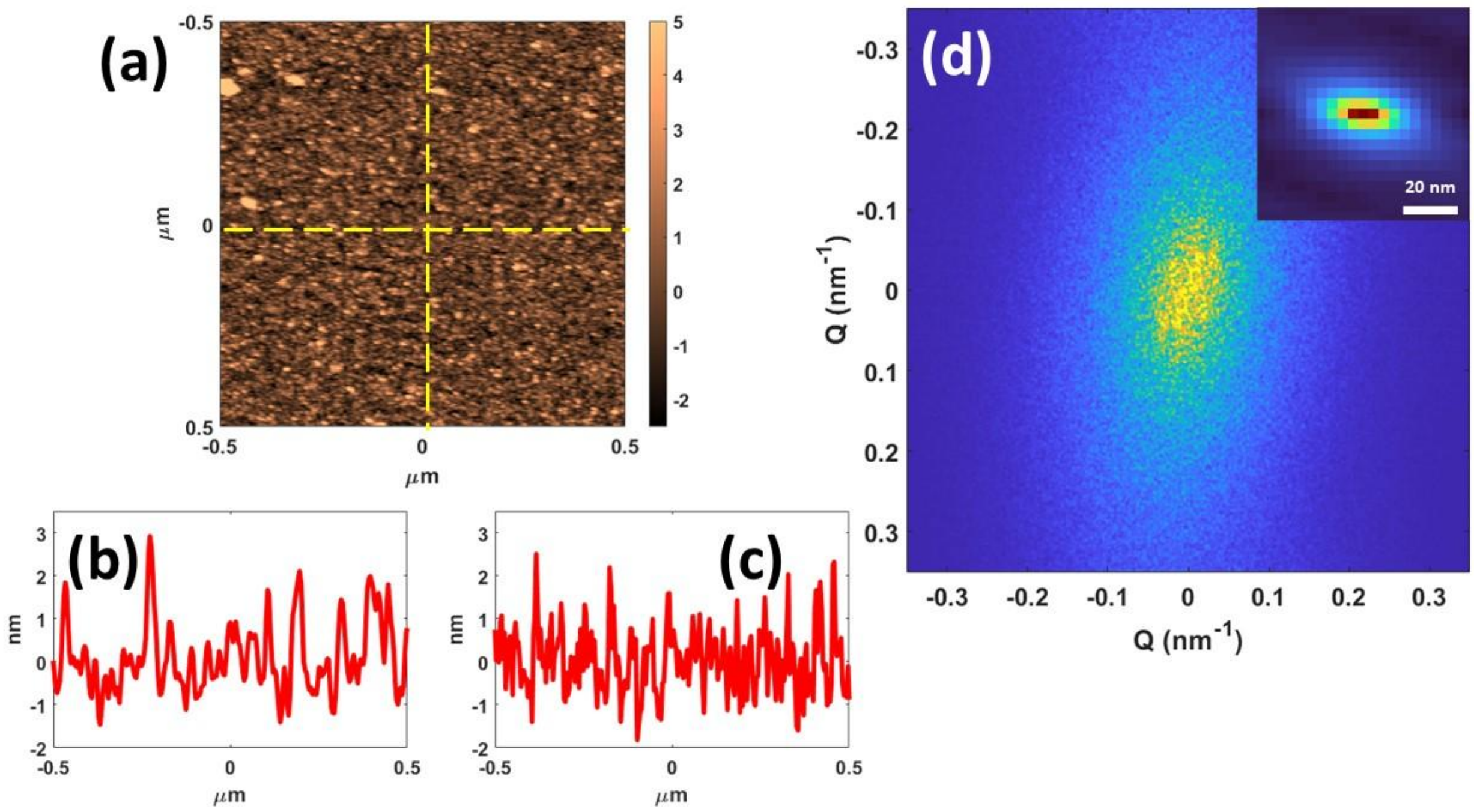

**Figure S15. (a)** 1x1 µm$^2$ AFM image of $Si_3N_4$/Pt(50 nm) sample. **(b)** Horizontal and **(c)** vertical profiles taken along the dashed yellow lines in panel **(a)**. Statistical analysis of these height profiles yields a $\sigma_R$ of 1.1 nm. **(d)** shows the Fourier transform of the AFM image. The top-right inset provides the central part of the autocorrelation function, revealing an anisotropic correlation length of 12±2 nm in the horizontal direction and 8±2 nm in the vertical direction.

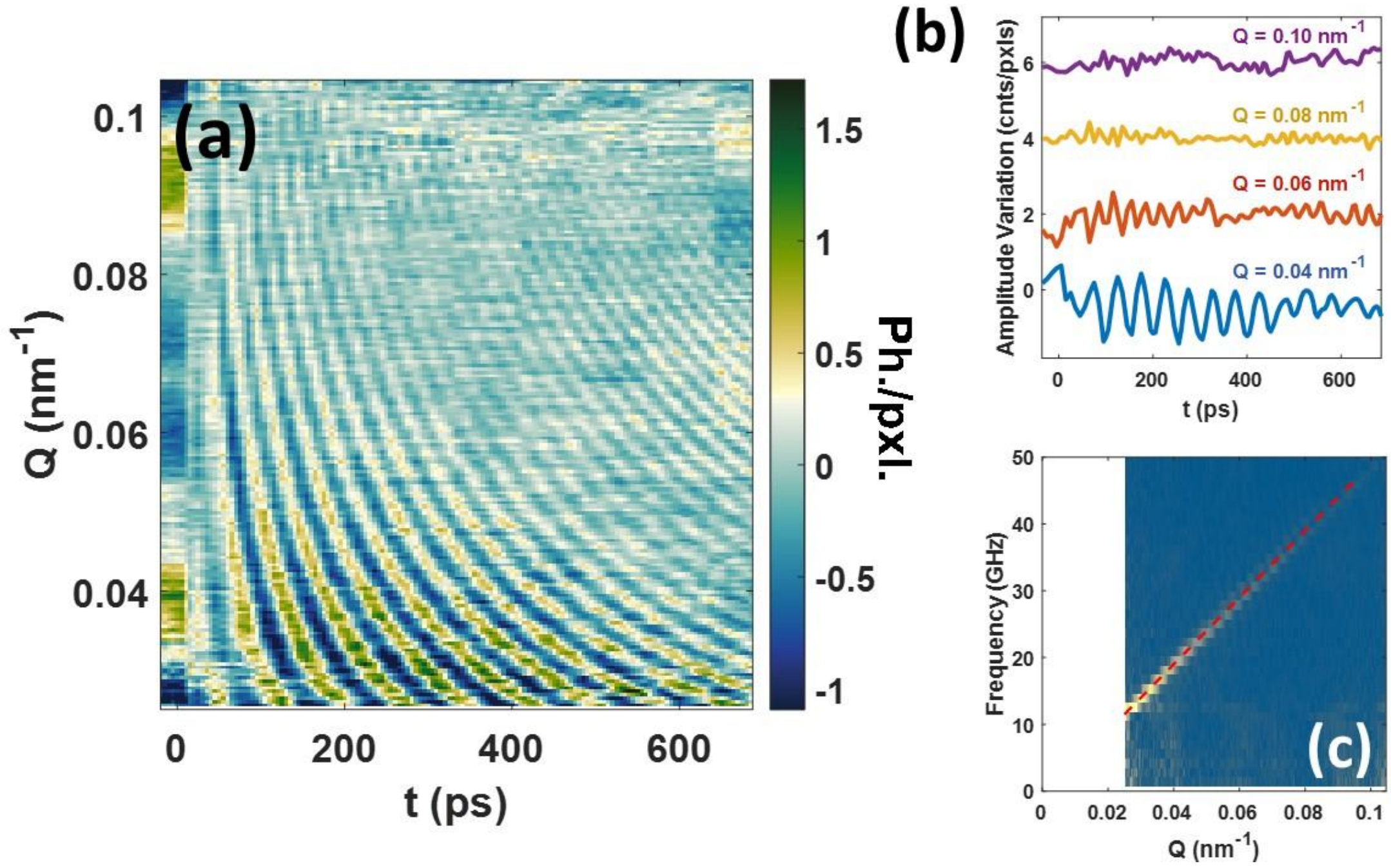

**Figure S16.** Radially averaged differential diffuse scattering intensity $S(Q,t)$ as a function of $Q$ and $t$ for the GaAs bulk sample. **(b)** Time dependence traces at selected values of $Q$ in the range 0.02 – 0.10 nm$^{-1}$. **(c)** Fourier transform of the data in panel **(a),** with the dashed red line indicating the linear dispersion of the first Rayleigh mode (same as Fig. 3(c) of main text).

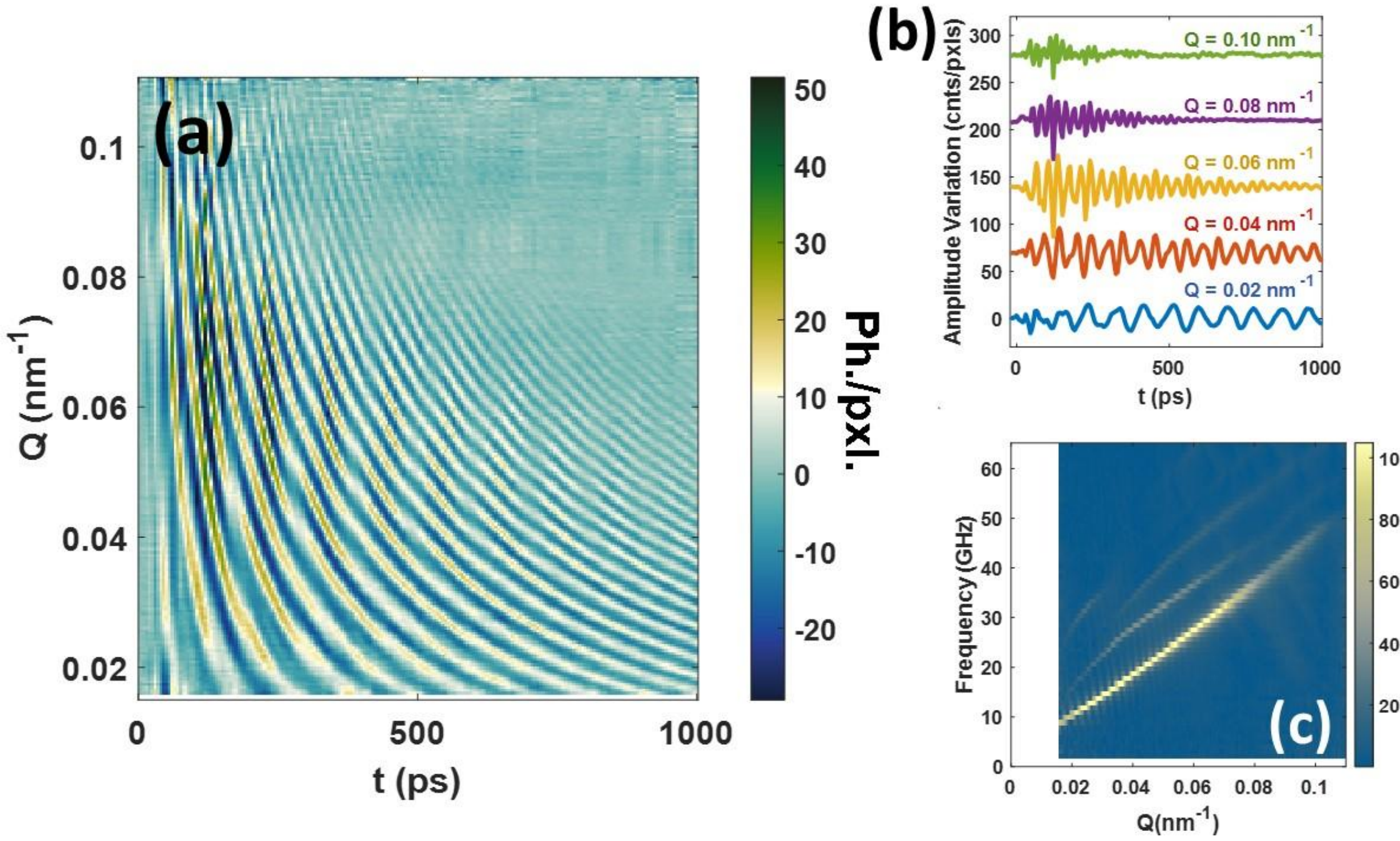

**Figure S17. (a)** Radially averaged differential diffuse scattering intensity $S(Q,t)$ as a function of $Q$ and $t$ for the Si/Ti(100 nm) sample. **(b)** Time dependence traces at selected values of $Q$ in the range 0.02 – 0.10 nm$^{-1}$. **(c)** Fourier transform of the data in panel **(a)** (same as Fig. 3(d) of main text).

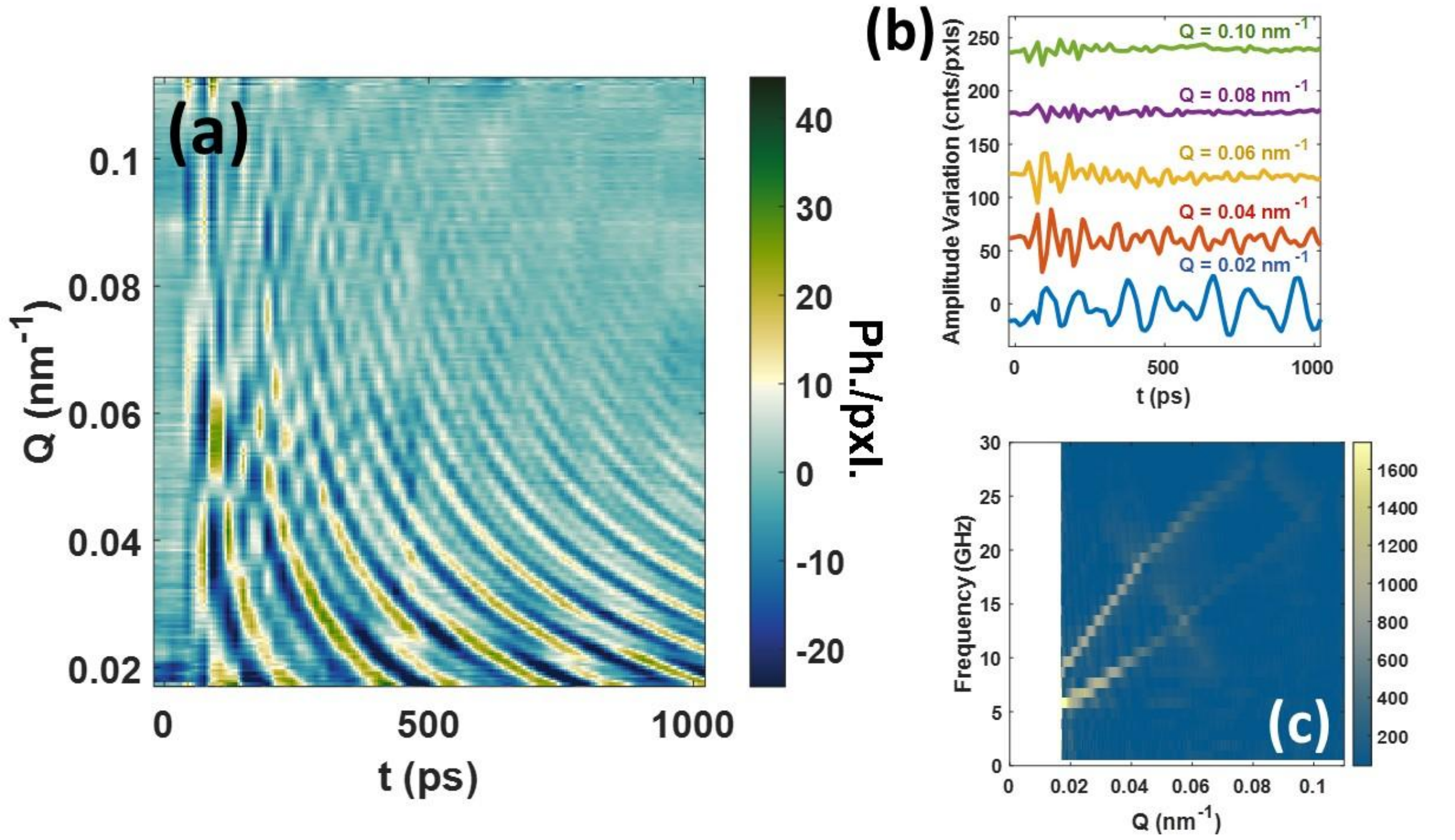

**Figure S18. (a)** Radially averaged differential diffuse scattering intensity $S(Q,t)$ as a function of $Q$ and $t$ for the Si/Pt(50 nm) sample. **(b)** Time dependence traces at selected values of $Q$ in the range 0.02 – 0.10 $nm^{-1}$. **(c)** Fourier transform of the data in panel **(a)**.

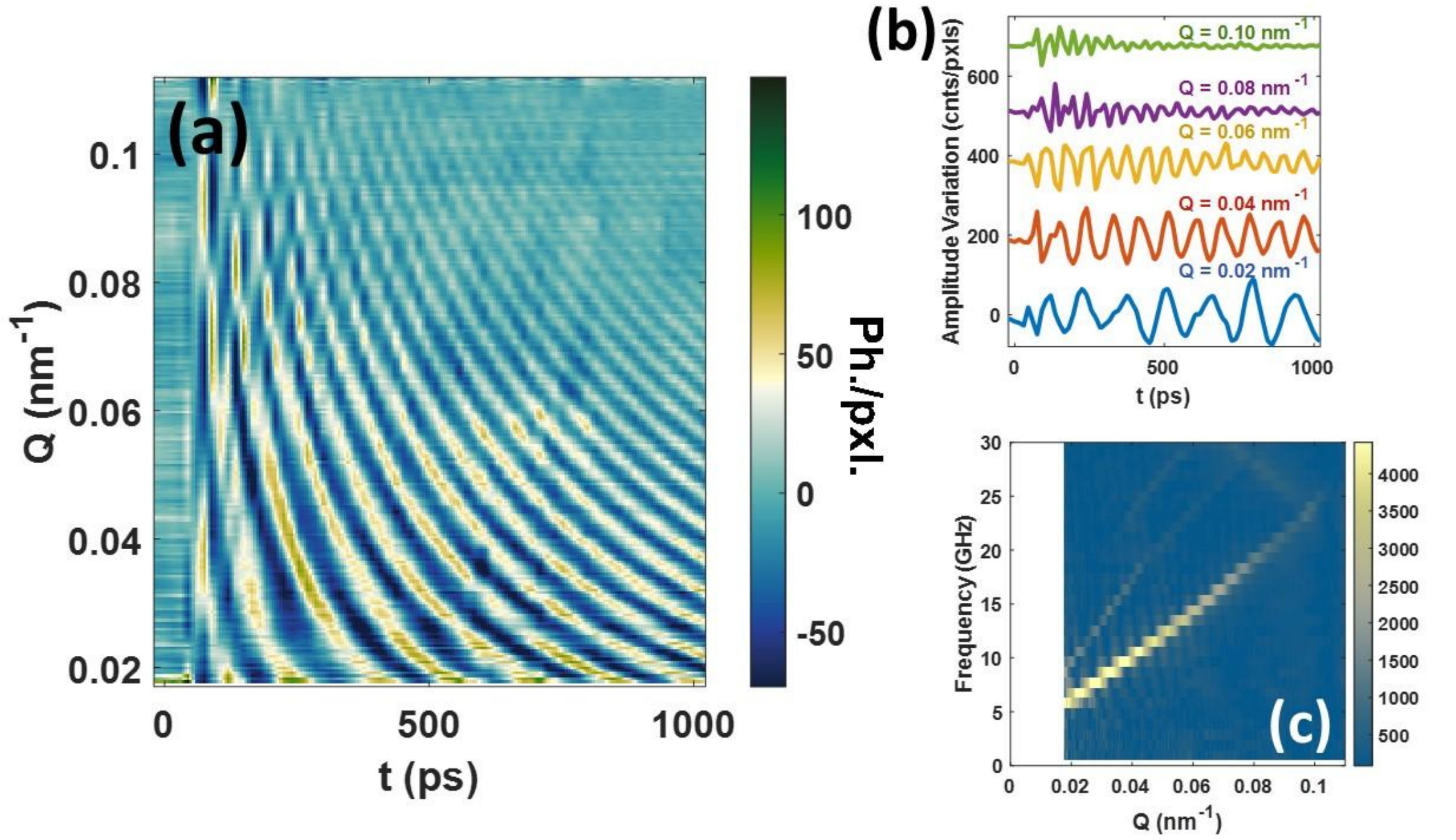

**Figure S19. (a)** Radially averaged differential diffuse scattering intensity $S(Q,t)$ as a function of $Q$ and $t$ for the $Si/Si_3N_4/Pt$(50 nm) sample. **(b)** Time dependence traces at selected values of $Q$ in the range 0.02 – 0.10 $nm^{-1}$. **(c)** Fourier transform of the data in panel **(a)**.

# S2. Impact of SAW propagation anisotropy in DES scattering.

The effects of elastic anisotropy on acoustic wave propagation in cubic crystals, and their impact on deposited amorphous films, have been extensively studied in the past using both surface Brillouin scattering [40] and optical transient grating [41]. It is well known that besides the SAW mode (refereed to anisotropic media as the generalized surface wave, GSW), an elastically anisotropic half space may support pseudo-surface acoustic wave (PSAW). GSWs are true surface-bound solution analog of Rayleigh waves, i.e. their amplitude decays exponentially into the bulk and they do not radiate energy away, while PSAWs are not fully bound states, their velocity is higher than that of bulk shear modes for a given crystal direction and they couple to bulk modes leaking energy as they travel. The angular dependence of these different modes presents a specific behavior for (001) oriented crystal. Along the [100] directions GSWs motion is on the sagittal plane: a pure Rayleigh wave. For finite angles with respect to the [100] directions a shear horizontal component appears, turning into a pure horizontally polarized bulk shear wave along the [110] direction. At the same time along the [110] direction the PSAW turns from a leaky mode into a pure surface mode with slightly higher group velocity compared to GSW [40,42].

Figure S20 shows the effect of elastic anisotropy on DES scattering of (001) GaAs crystal. In the employed scattering geometry, the crystallographic [110] directions are along the X and Y axis of the detector CCD plane. By restricting the data analysis along these two different crystallographic directions (light gray shaded regions in Figs. S20(a) and S20(b)), we observe a distinct difference in the dispersion curves of surface acoustic phonons (Fig. S20(c)). In agreement with previously reported results [40,43], the velocity of PSAW along the [110] direction (3080 m/s) is ~ 4% higher than for the [100] (2950 m/s). The slightly higher velocities found with DES approach with respect to literature values [40,43] can be due to an imprecision of a few mm in the evaluation of the sample to detector distance.

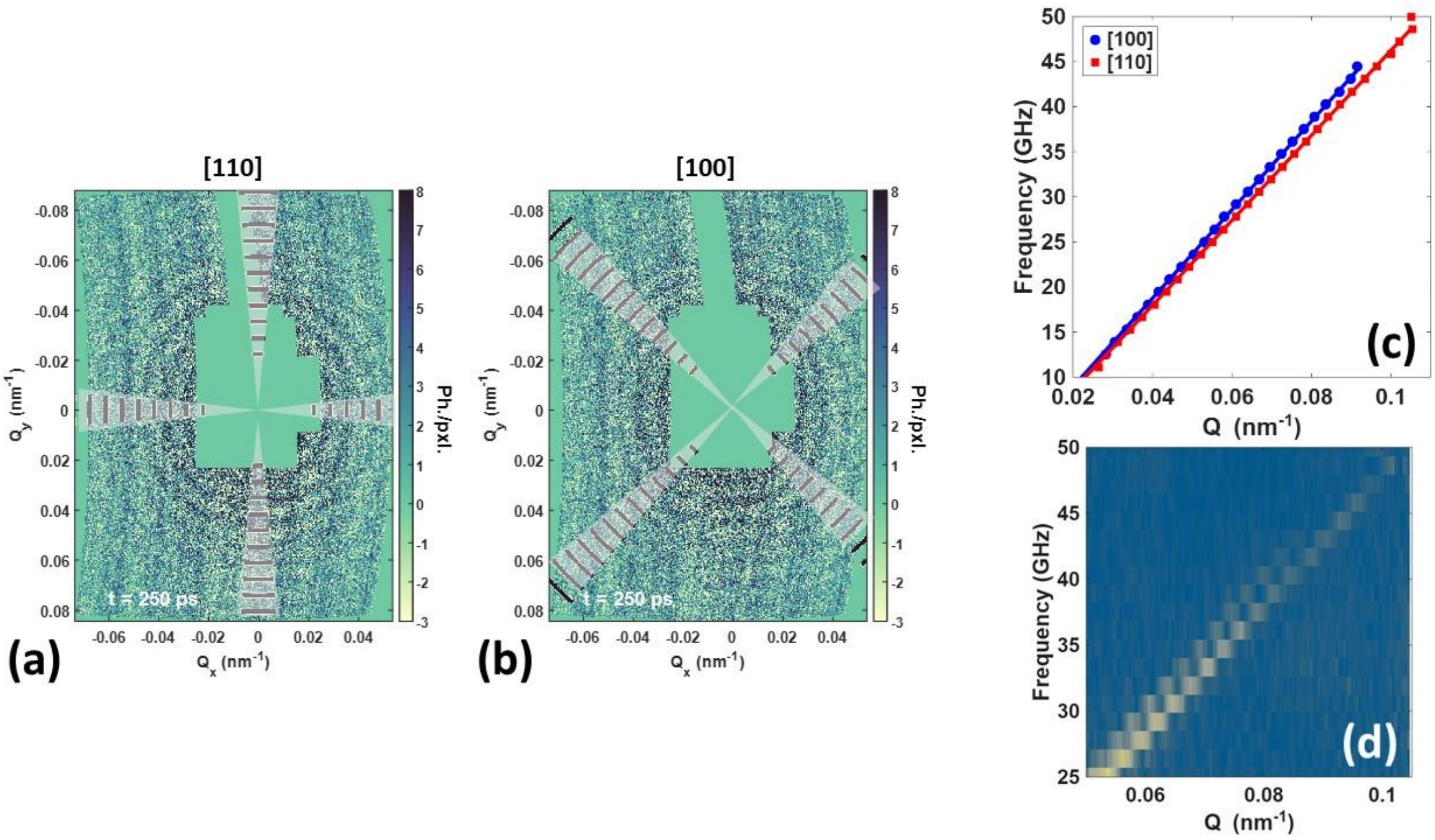

Figure S20 (a)-(b) Fringe patterns in the $\Delta I_{DES}(Q_x, Q_y, t)$ for (001) GaAs bulk sample at $t = 250$ ps after the optical excitation. Light gray shaded areas indicate the ±5° sectors used to perform the angular average as a function of Q near the [100] and [110] directions. (c) Dispersion relations (symbols) for the two crystallographic directions obtained from the maxima of Fourier transform of $S(Q, t)$ map calculated using shaded areas in panel (a) and (b). The SAW velocity is consistent with 3080 m/s and 2950 m/s, for the [110] and [100], respectively. (d) Magnified view of the Fourier transform of $S(Q, t)$ map reported in Figure 3(c) of main text.

Interestingly, evidence of the SAW velocity anisotropy is also observed in the Fourier transform map based on the full azimuthal average. A magnified view of the Fourier transform of $S(Q,t)$ reported in Fig. 3(c) of the main text reveals a splitting at $Q \gtrsim 0.06$ nm$^{-1}$ (Fig. S20(d)). This splitting can be attributed to the larger contribution to the Fourier transform intensity from the nearly flat regions of the angular phase velocity dependence in the vicinity of the [100] and [110] directions [40].
Similar behavior is expected also for Si and Ge, however, the lower signal to noise ratio for these samples (Figs. S8 and S10) does not allow us to evaluate the SAW velocity along specific crystallographic directions.
The effects of elastic anisotropy on acoustic wave propagation could influence also the SAW velocity for polycrystalline metallic films deposited on Si (001) substrates. However, numerical calculations for a Ti film on a Si (100) substrate shows that these effects become negligible when the product of the wavevector $Q$ and film thickness $d$ exceeds 2 [41]. Figures S21(a)-(d) present the phonon dispersion curves along two crystallographic directions for Si/Ti(100 nm), Si/Pt(50 nm), Si/Co(75nm)/Ta(2 nm) and Si/Ta(75nm) samples, respectively, integrating the $\Delta I_{\mathrm{DES}}\left(Q_x, Q_y, t\right)$ signals along the [100] and [110] directions. No evidence of anisotropy between the two directions induced by the Si substrate is observed in any of the samples within the investigated $Q$-range, confirming that substrate elastic anisotropy has a negligible effect on the SAWs propagation when the phonon wavelength is comparable to the layer thickness [41].

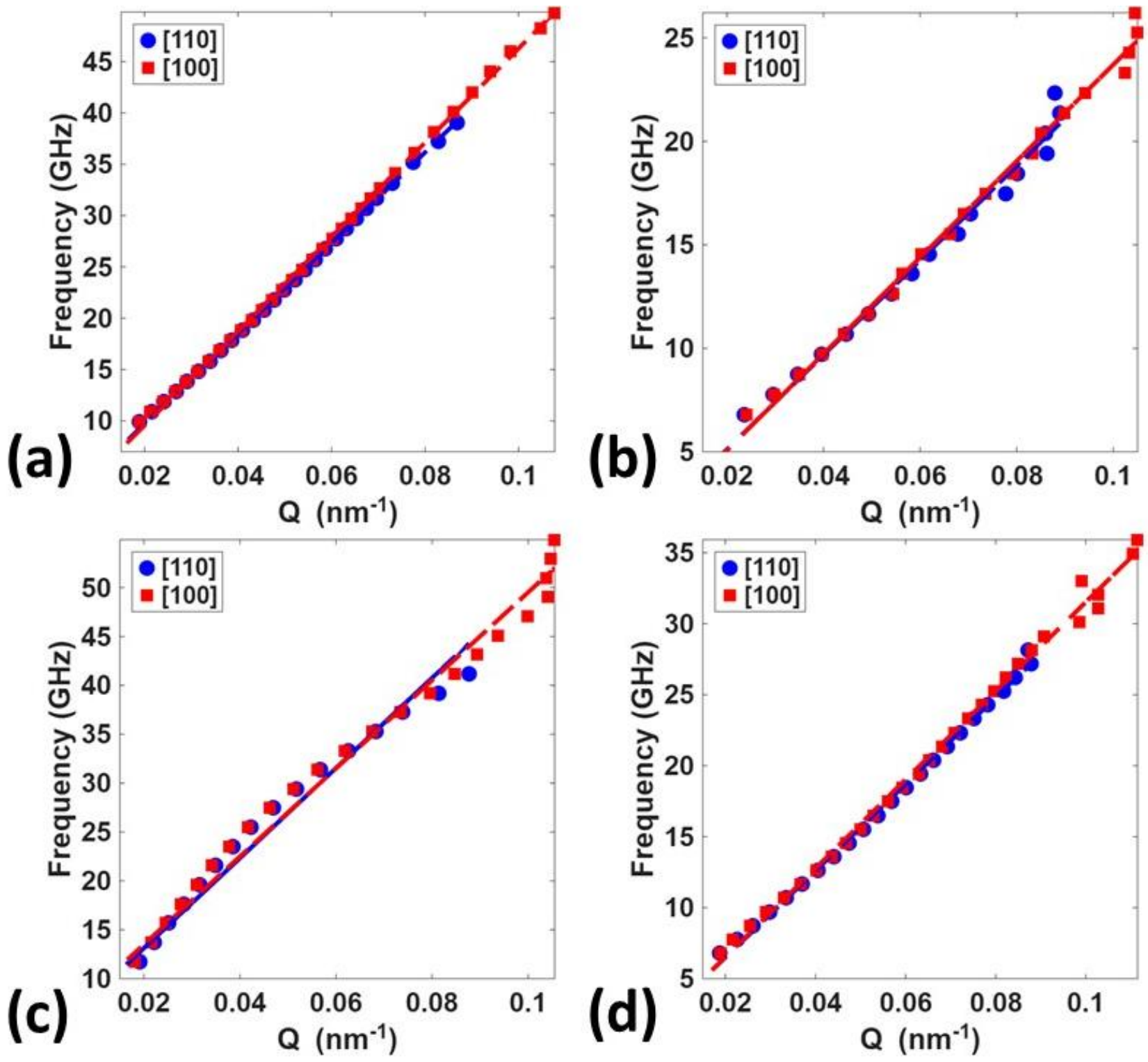

Figure S21 Dispersion relations of the first Rayleigh mode along the [110] (blue) and [100] (red) crystallographic directions obtained from the maxima of the Fourier transform of $S(\mathrm{Q},\mathrm{t})$ map for: (a) Si/Ti(100 nm), (b) Si/Pt(50 nm), (c) Si/Co(75 nm)/Ta(2 nm), (d) Si/Ta(75 nm). Blue and red dotted lines are guides for the eye.

# S3. 790 nm vs 395 nm excitation

Figure S22 presents a comparative analysis of DES scattering from a [Pt(4 nm)/Al(4 nm)]×4 multilayer stack fabricated on a Si/$SiO_2$ substrate, subjected to optical excitation at two distinct wavelengths: 790 nm and 395 nm. In both cases, the incident energy density at the sample surface was maintained at 17 ± 3 mJ/cm². The laser pulse was *p*-polarized at 790 nm and *s*-polarized at 395 nm. Figures S22(a) and S22(b) display the differential DES images recorded on the CCD detector. The fringe patterns are similar for both excitation wavelengths, indicating that the underlying physical mechanism is largely independent of the optical pumping wavelengths and polarization. The corresponding $S(Q, \Delta t)$ maps are shown in Figures S22(c) and S22(d), while Figures S22(e) and S22(f) present representative line cuts at selected wave vectors in the range of 0.04 to 0.10 nm⁻¹. From these data, the extracted surface phonon propagation velocities are 1490 ± 30 m/s for 790 nm excitation and 1520 ± 30 m/s for 395 nm excitation. The excellent agreement between the data underscores the generality and robustness of the observed phenomenon.

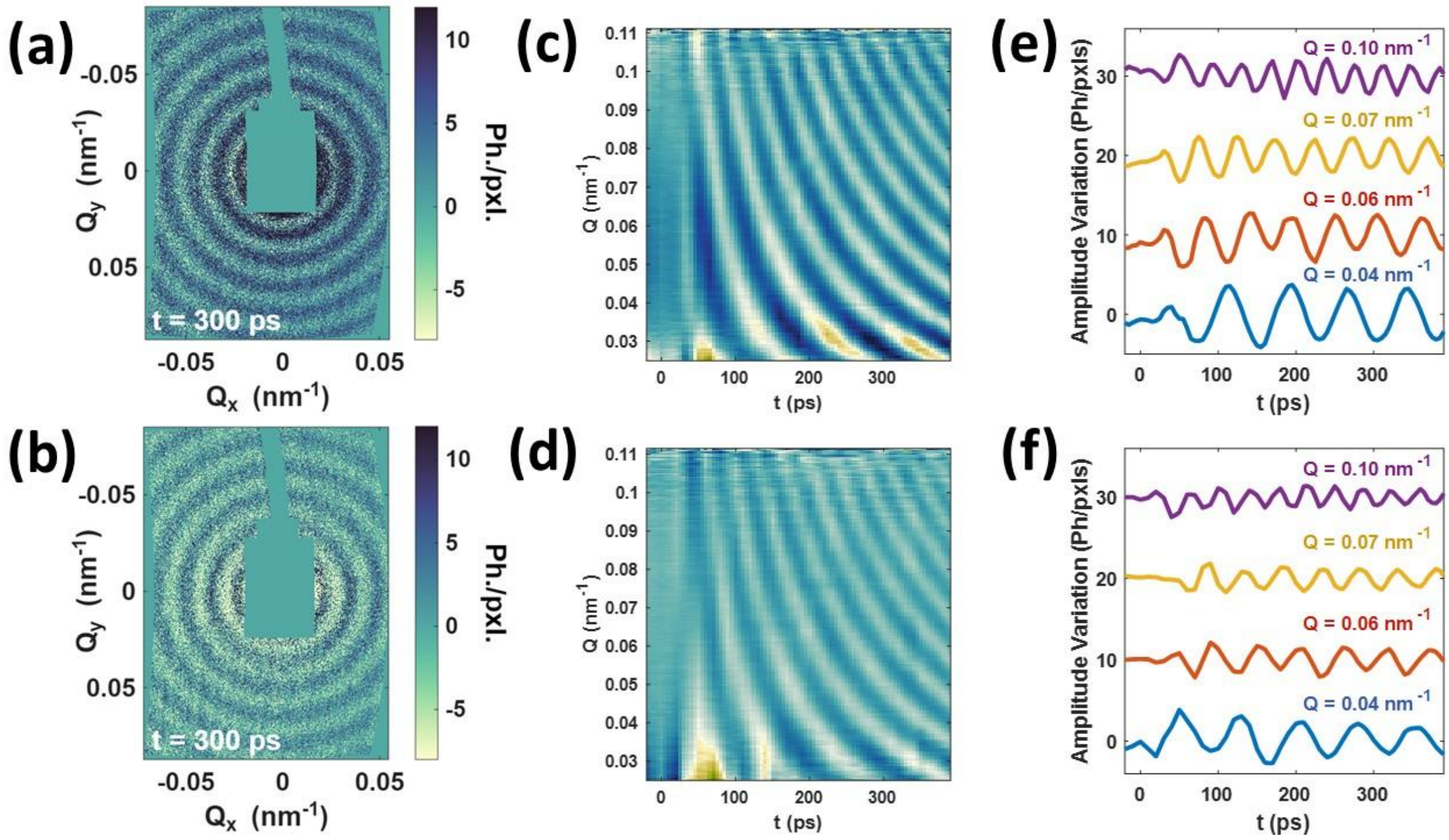

**Figure S22. (a)** and **(b)** Differential DES images, $\Delta I_{EDS}(Q_x, Q_y, t = 300\,ps)$, on [Pt(4nm)/Al(4nm)]×4 multilayer stack grown on Si/$SiO_2$ substrate for 790 nm and 395 nm optical excitation respectively. **(c)** and **(d)** Radially averaged differential diffuse scattering intensity $S(Q, t)$ as a function of $Q$ and $t$ for the two optical excitation wavelengths. **(e)** and **(f)** Selected line profiles at specific values of $Q$ in the range 0.04 – 0.10 nm$^{-1}$ for 790 nm and 395 nm pumping respectively.

# S4. Additional evidence of DES dependence on surface roughness.

This section provides additional support for the influence of surface roughness on the amplitude of the fringe patterns observed in diffuse DES scattering. A multilayer [Ta(2.4nm)/Pt(2.4nm)]×4 stack was deposited both on Si and on $Si/Si_3N_4$(100 nm) substrates. The AFM reveals substantial differences in surface morphology between the two samples. The sample grown on the Si substrate (Fig. S23(a)) exhibits a random structure with an RMS surface roughness of approximately 0.5 nm. In contrast, the sample grown on the $Si/Si_3N_4$ substrate shows terrace-like features with a significantly higher roughness of about 2.4 nm (Fig. S23(b)). As a result of this pronounced surface structuring, the static diffuse scattering from the multilayer on the $Si/Si_3N_4$ substrate exhibits a broad maximum centered around $Q_x = 0.05$ nm$^{-1}$ (Fig. S23(d)), a feature absents in the scattering pattern of the sample grown on bare Si (Fig. S23(c)). Under identical optical excitation conditions, the amplitude of the differential DES signal $\Delta I_{\mathrm{DES}}(Q_x, Q_y, t = 300\ ps)$ is over an order of magnitude greater for the sample on $Si/Si_3N_4$ substrate (Fig. S23(f)) compared to the smoother Si-grown sample (Fig. S23(e)). In agreement with Eq.(2) of the main text, further evidence for the heterodyne enhancement of the surface acoustic wave signal by surface roughness is provided in Figure S23(f). Specifically, the maximum amplitude of the differential signal $\Delta I_{\mathrm{DES}}(Q_x, Q_y)$ occurs near $Q_x = 0.05\ nm^{-1}$, coinciding with the momentum transfer where the static diffuse signal from the rough morphology reaches its maximum (Fig. S23(d)).

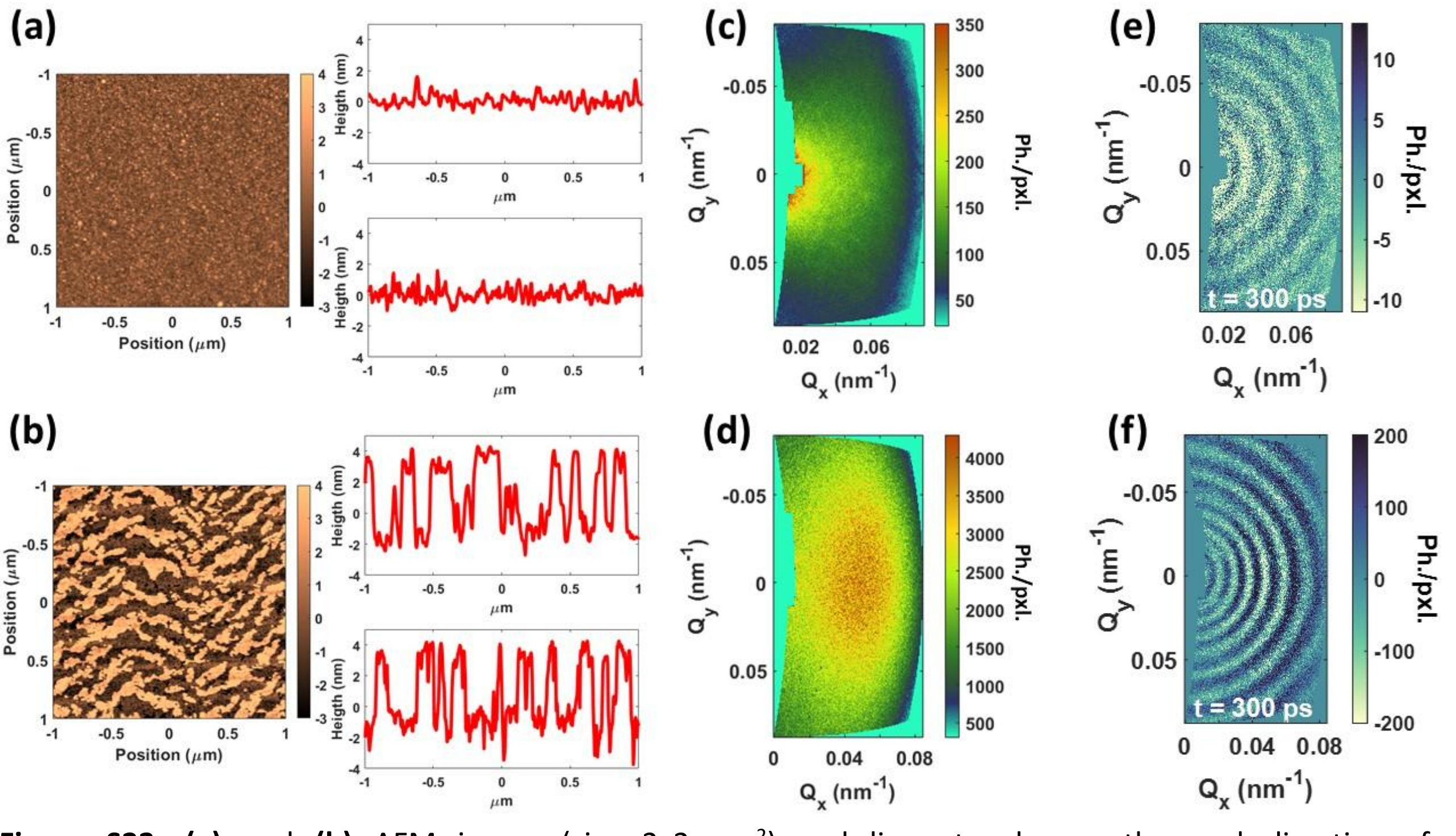

**Figure S23. (a)** and **(b)** AFM image (size 2x2 μm$^2$) and lineouts along orthogonal directions for [Ta(2.4nm)/Pt(2.4nm)]$_4$ multilayer stack grown on Si and $Si/Si_3N_4$(100 nm) substrate respectively. Statistical analysis of these height profiles for the multilayer structure grown on Si substrate yields a surface roughness $\sigma_R$ of about 0.5 nm, while for the sample grown of $Si/Si_3N_4$(100 nm) the roughness is of about 2.4 nm. **(c)** and **(d)** Static diffuse scattering collected with the specular reflected beam placed on the left side of the detector chip, for the two sample. **(e)** and **(f)** Differential DES images, $\Delta I_{\mathrm{DES}}(Q_x, Q_y, t)$, for the two cases recorded at $t = 300\ ps$ at a pump fluence of 20 mJ/cm$^2$.

# S5. Surface phonon dispersion calculations.

In the wavelength range down to tens of nm, the surface phonon dispersion can be calculated within the framework of continuum elasticity. The methodology for such calculations for layer/substrate systems are well known [44]. We are using a modified approach based on surface Green's functions: instead of looking for zeros of the boundary conditions determinant [44], we are looking for singularities of surface Green's functions [43,45], which avoids spurious solutions at phase velocities equal to bulk acoustic velocities of the layers. A matrix transfer formalism is employed to facilitate the calculations for systems with a large number of layers [46]. Table S2 lists the properties of the layer materials used for the calculations of SAW dispersion relations reported in Fig. 2(c) and Fig. 3(d). For the Si substrate, the following values of the density and elastic constants were used: $\rho$ = 2330 kg/m$^3$, $C_{11}$ = 166 GPa, $C_{12}$ = 64 GPa, $C_{44}$=79.6 GPa. The calculations were done for the wave vector direction along the [110] direction of the (001) Si substrate; since we did not observe any noticeable anisotropy of the surface phonon velocity (note the perfectly circular shape of the fringes in Fig. 1(e-f)), the choice of the propagation direction within the (001) plane was not important.

| Material | Density (kg/m$^3$) | Longitudinal velocity (m/s) | Transverse velocity (m/s) |
|---|---|---|---|
| Al | 2699 | 6444 | 3109 |
| Pt | 21500 | 4136 | 1721 |
| Ti | 4506 | 6052 | 3102 |
| amorphous $SiO_2$ | 2200 | 5884 | 3729 |

**Table S2.** Material properties used for surface phonon dispersion calculations.